\documentclass[aps,prl,preprint,superscriptaddress]{revtex4-1}
\usepackage{amsmath}
\usepackage{amsfonts}
\usepackage{bm}
\usepackage{graphicx}
\usepackage{hyperref}
\usepackage{pifont}
\usepackage{color}
\usepackage{tabularx}

\begin{document}

\title{Valley Zeeman effect and Landau levels in Two-Dimensional Transition Metal Dichalcogenides}

\author{Fengyuan Xuan}
\affiliation{Centre for Advanced 2D Materials and Graphene Research Centre, National University of Singapore, Block S16, Level 6, 6 Science Drive 2, Singapore 117546}
\author{Su Ying Quek}
 \email{phyqsy@nus.edu.sg}
\affiliation{Department of Physics, Centre for Advanced 2D Materials and Graphene Research Centre, National University of Singapore, 6 Science Drive 2, Singapore 117546, Singapore.}

\date{\today}

\begin{abstract}
This paper presents a theoretical description of both the valley Zeeman effect (g-factors) and Landau levels 
in two-dimensional H-phase transition metal dichalcogenides (TMDs) using the Luttinger-Kohn approximation with spin-orbit coupling. 
At the valley extrema in TMDs, energy bands split into Landau levels with a Zeeman shift in the presence of a uniform out-of-plane external magnetic field.
The Landau level indices are symmetric in the $K$ and $K'$ valleys. 
We develop a numerical approach to compute the single band g-factors from first principles without the need for a sum over unoccupied bands. 
Many-body effects are included perturbatively within the GW approximation. 
Non-local exchange and correlation self-energy effects in the GW calculations increase the magnitude of single band g-factors compared to those obtained from density functional theory. 
Our first principles results give spin- and valley-split Landau levels, in agreement with recent optical experiments. 
The exciton g-factors deduced in this work are also in good agreement with experiment for the bright and dark excitons in monolayer WSe$_2$, 
as well as the lowest-energy bright excitons in MoSe$_2$-WSe$_2$ heterobilayers with different twist angles. 
\end{abstract}

\maketitle
\section{Introduction}
When a weak external magnetic field is applied to a periodic system, there are two physical effects on the band extrema. 
Firstly, Bloch states are reorganized into highly degenerate Landau Levels (LLs). 
Secondly, LLs are further shifted in energy, known as the Zeeman effect. 
In recent years, the seminal prediction of spin-valley coupling caused by broken inversion symmetry in semiconducting monolayer (ML) 
H-phase transition metal dichalcogenides (TMDs) \cite{Xiao2012} has led to a surge in theoretical and experimental interest 
in the response of TMDs to external magnetic fields for valleytronic applications \cite{Nanoph,Wang2015,NPJ2018,Wang2018,XX2016,NatPhys2015,Mak}. 
Inversion symmetry breaking and three-fold rotational symmetry also lead to valley-dependent optical selection rules \cite{Niu2008,RMP2010,Xiao2012,NC2012}, 
which have enabled experimentalists to measure the valley-dependent Zeeman effect \cite{NatPhys2015} and Landau levels \cite{Mak} in these TMDs. 
Theoretically, how energy band edges at each valley respond to the external magnetic field is essential to understand both phenomena. However,
most theoretical works target each effect separately.

On the one hand, the band edge Zeeman shift has been explained by the band- and valley-dependent orbital magnetic moment $m_{n\mathbf{k}}$, 
or Laud\'e g-factor $g^{\text{orb}}_{n\mathbf{k}}$, given by $m_{n\mathbf{k}}=g^{\text{orb}}_{n\mathbf{k}}\mu_B$.
In Ref. \cite{2d2015}, the authors show that Peierls substitution into a multi-band $\mathbf{k} \cdot \mathbf{p}$ Hamiltonian matrix results in a term that is linear in the external magnetic field,
which was used to define the Laud\'e g-factor.
However, the final expressions do not contain information on LLs.
Another approach that has been widely used is to define the orbital magnetic moment using semiclassical considerations \cite{Niu1996}. This semiclassical method also does not lead to expressions for LLs.
On the other hand, LLs in TMDs have been derived using the massive Dirac fermion model \cite{LL2013,LLPRB2013,LLPRL2013},
obtained by Peierls substitution into a two-band tight binding (TB) model. 
This model results in LL indices that are asymmetric between the $K$ and $K'$ valleys \cite{LL2013,LLPRB2013,LLPRL2013},
with no Zeeman term.

To explain experiments in which both the valley Zeeman effect and LLs are important \cite{Mak,LL2016}, 
one needs to find a general approach to take into account the two effects on an equal-footing framework.
However, in many cases, the above theories are patched together in an \textit{ad hoc} fashion to interpret 
transport and optical measurements in the literature \cite{LL2016,LL2017,Mak,LL2019}.
Specifically, the LLs obtained from the massive Dirac fermion model are further shifted by an additive Zeeman term, 
which is obtained from one of the models that deduced an expression for the Zeeman term without the LL effect. Such a simple mixing of different models is not convincing.
One objection is from Niu et al., who have suggested that the LLs from the massive Dirac fermion model should not be further shifted by any Zeeman term \cite{LLPRB2013}. 
Fundamentally, this confusion of how to treat both the Zeeman effect and LLs arises from the fact that the Zeeman effect and LLs have been explained separately 
with different model Hamiltonians. The use of different Hamiltonians also leads to controversies in the interpretation of the g-factors.

Existing quantitative predictions of the single band g-factor can be summarized into two approaches \cite{Wang2018}.
The first approach is based on a phenomenological model \cite{2d2019}, where the orbital magnetic moment is partitioned into atomic and valley terms, 
so that the total single band g-factor consists of three terms, 
namely the spin, atomic and valley terms \cite{Nanoph}. 
Since the valence band maximum (VBM) and conduction band minimum (CBM) of TMDs are composed mainly of $d$ orbitals, 
atomic contributions are taken as their magnetic quantum numbers $\pm 2$ and $0$ \cite{atomPRB2018}.
The valley term is approximated to be inversely proportional to the effective mass \cite{XD2007,Niu2008,PRL2014,Mak}. 
However, for models with more than two bands, such a relation does not hold anymore \cite{2d2015}.
Another problem is that the partitioning of the orbital magnetic moments into atomic and valley terms has not been justified properly - in particular, 
it is not clear whether they are additive and if there are other contributions left.
The second approach is based on Peierls substitution into an effective multi-band $\mathbf{k} \cdot \mathbf{p}$ Hamiltonian \cite{2d2015}, which has been discussed above.
Contrary to the phenomelogical model, the authors argue that atomic terms are vanishing and the effect from remote bands is believed to be important. 
In a later work, it is shown that Hamiltonians with different number of bands lead to very different exciton g-factors \cite{kpPRB2017}.
Although both approaches can give g-factors in reasonable agreement with experiment, their numerics very much depend on the specific model Hamiltonian chosen.

Korm\'{a}nyos et al. have proposed to reduce a multi-band $\mathbf{k} \cdot \mathbf{p}$ Hamiltonian to a single-band model using L\"{o}wdin-partitioning \cite{NewJPhys}. 
In this case, the energy expressions for the band extrema contain both the Zeeman term and LLs.
However, an apparent drawback of this approach is that the g-factor used in the Zeeman term is model-dependent, and in particular, 
also does not have the atomic contribution that was suggested in the phenomelogical models.
Thus, given all the above considerations, it is unsatisfactory to rely on existing models to obtain an accurate description of the valley Zeeman effect and LLs in TMDs.

In this work, to describe both the valley Zeeman effect and LLs in two-dimensional (2D) TMDs, 
we propose using a general Hamiltonian including spin-orbit coupling (SOC) for an electron in a periodic potential perturbed by a uniform external magnetic field. 
Our key results are as follows: within the Luttinger-Kohn (LK) approximation \cite{LK1955},
the valley band edges split into LLs with a valley-dependent Zeeman shift.  
In contrast to previous predictions \cite{LLPRB2013,NewJPhys,Mak,2d2019,2d2015}, LL indices obtained from the present work are symmetric with respect to the $K$ and $K'$ valleys, 
and the orbital magnetic moment describing the Zeeman shift is shown to be equivalent to the compact Berry-curvature-like expression derived using the semiclassical approach \cite{Niu1996}.
We further extend the LK approximation to electrons in a non-local periodic potential, where the compact form of the magnetic moment remains unchanged.
This extension allows us to include non-local exchange and correlation effects in our predictions for the single band g-factor, which have not been discussed before.
We implement an approach to evaluate the Berry-curvature-like expression for the g-factor without the need for a sum over unoccupied bands, using first principles Hamiltonians. 
We use density functional theory (DFT) as well as many-body perturbation theory within the GW approximation, and obtain results for prototypical TMD systems. 
Firstly, the single band g-factor and Berry curvature obtained using DFT are found to be quite different from those derived using TB Hamiltonians.
Deriving analytically separate components of the orbital magnetic moment using a tight-binding basis, we provide clear definitions of
the valley and atomic terms of the orbital magnetic moment and furthermore, uncover an additional cross term,
which arises from a coupling between the phase winding of Bloch states and the parent atomic orbitals.
The deviation of the TB results from DFT originates from the omission of atomic and cross terms.
Secondly, compared to DFT, GW self-energy effects increase the magnitude of single band g-factors significantly while keeping exciton g-factors in ML WSe$_2$ approximately unchanged.
For the interlayer excitons in MoSe$_2$-WSe$_2$ heterostructures, GW results agree better with experiments than DFT.
Finally, the single band g-factors, together with our predictions for the LL spacings, result in spin- and valley-split LLs, consistent with the optical experiment by Mak et al. \cite{Mak}.

Our theoretical formalism is presented in Section (II) and in Appendix \ref{AppA}-\ref{AppC}, where treatment of non-local potentials is specifically discussed. 
Our numerical method is also presented in Section (II), with numerical results shown and discussed in section (III).  
In Section (IV), we summarize the paper and discuss briefly the possibility of accounting for the energy-dependence of the self-energy.
Derivations of the atomic, valley and cross terms of the orbital magnetic moment are included in Appendix \ref{AppD}.

\section{Theoretical Approach}
\subsection{Formalism}
We start with the general Hamiltonian of an electron in a local periodic potential with SOC, 
in the presence of an external magnetic field $\mathbf{B}=\nabla \times \mathbf{A}=(0,0,B_z):$
\begin{eqnarray}
\begin{split}
&{\mathbb{H}}=\frac{(\mathbf{p}+e\mathbf{A})^2}{2m_e}+V(\mathbf{r})+\mathbb{H}^{SOC}+\mathbb{H}^{spin}\\
&\mathbb{H}^{SOC}=\frac{\hbar}{4m_e^2c^2}(\bm{\sigma}\times \nabla V)\cdot (\mathbf{p}+e\mathbf{A})\\
&\mathbb{H}^{spin}= g_s \frac{1}{2}\bm{\sigma} \cdot \mathbf{B}
\label{HB}
\end{split}
\end{eqnarray}
where $m_e$ is the electron mass, $V(\mathbf{r})$ is the local periodic potential, $\mu_B$ is the Bohr magneton, and $g_s \approx 2$ is the free electron g-factor. 
$\bm{\sigma}$ refers to the Pauli matrices.
Because $s_z$ is a good quantum number at $K$ and $K'$ in the H-phase TMDs, 
the spin diagonal components of Eq.~(\ref{HB}) give the solutions for spin up and down states. 
Henceforth, the spin Zeeman term $\mathbb{H}^{spin}$ is dropped from our equations, and will be added on again in the final result for the energy levels.

We first consider the energy levels of the Hamiltonian of ML TMD without the magnetic field. 
Using a second order expansion of the energy levels around $K$, the non-degenerate quasiparticle energy levels $E_{n}(\mathbf{K+q})$ are given by (see Appendix \ref{AppA})
\begin{eqnarray}
E_{n\mathbf{K}+\mathbf{q}} = E_{n\mathbf{K}} +\frac{\hbar^2 \mathbf{q}^2}{2m_e} + E_{n\mathbf{K}}^{(2)} (\mathbf{q})
\label{k-dot-p}
\end{eqnarray}
where
\begin{eqnarray}
E_{n\mathbf{K}}^{(2)}(\mathbf{q}) = (\frac{\hbar}{m_e})^2 \sum_{m \neq n} \frac{|<u_{n\mathbf{K}}|\mathbf{q} \cdot \bm{\pi}|u_{m\mathbf{K}}>|^2}{E_{n\mathbf{K}}-E_{m\mathbf{K}}}
\label{E2}
\end{eqnarray}
Here, $\mathbf{q}$ is the crystal momentum and $u_{n\mathbf{K}}$ are the periodic part of the Bloch eigenstates of the Hamiltonian without any magnetic field. 
Note that if SOC is neglected, $\bm{\pi}=\mathbf{p}+\frac{\hbar}{4m_e c^2}(\bm{\sigma}\times \nabla V)$ is simply replaced by $\mathbf{p}$ and all the formalism in what follows remains the same.
Luttinger and Kohn \cite{Lut1951,LK1955} have shown that if the magnetic field is treated as a perturbation, 
then, dropping $\mathbb{H}^{spin}$, the energy levels of the Hamiltonian in Eq.~(\ref{HB}) can be obtained by solving an effective eigenvalue problem:
\begin{eqnarray}
E_{n}(\mathbf{K+\hat{q}})\Psi_{\alpha}=\epsilon_{\alpha}\Psi_{\alpha}
\label{Lutt}
\end{eqnarray}
where $\mathbf{q}$ in $E_{n}(\mathbf{K}+\mathbf{q})$ is replaced by the operator $\hat{\mathbf{q}}=\mathbf{p}/\hbar+e\mathbf{A}/\hbar$.
The effective Hamiltonian $E_{n}(\mathbf{K+\hat{q}})$ required in Eq.~(\ref{Lutt}) can then be shown to be
\begin{eqnarray}
E_{n\mathbf{K}+\hat{\mathbf{q}}}=E_{n\mathbf{K}} +\frac{\hbar^2 (\hat{q}_x^2+\hat{q}_y^2)}{2m^*} - \mathbf{m}_{n\mathbf{K}} \cdot \mathbf{B}
\label{effH}
\end{eqnarray}
where the effective mass $ m^*$ is,
\begin{eqnarray}
\frac{1}{m^*} = \frac{1}{m_e}+\frac{1}{m_e^2} \sum_{m \neq n} \frac{2|\Pi^x_{nm}|^2}{E_{n\mathbf{K}}-E_{m\mathbf{K}}}
\end{eqnarray}
with the matrix elements $\bm{\Pi}_{nm} = <u_{n\mathbf{K}}|\bm{\pi}|u_{m\mathbf{K}}>$, 
and the orbital magnetic moment $\mathbf{m}_{n\mathbf{K}}$ is defined by coefficient of the linear term on the external magnetic field.
\begin{eqnarray}
\mathbf{m}_{n\mathbf{K}}=-i\frac{\mu_B}{m_e} \sum_{m \neq n} \frac{\bm{\Pi}_{nm}\times \bm{\Pi}_{mn}}{ E_{m\mathbf{K}}-E_{n\mathbf{K}} }
\label{defm}
\end{eqnarray}
Here, we have neglected terms involving $q_z$ and $\Pi^z_{nm}$ as there is no band dispersion along the $z$ direction in 2D materials. 
The numerator from Eq.~(\ref{E2}) results in a term of the form $\{\hat{q}_x , \hat{q}_y\} Re[\Pi^x_{nm}\Pi^y_{mn}]+[\hat{q}_x, \hat{q}_y] i 
Im[\Pi^x_{nm}\Pi^y_{mn}]$. Using $[\hat{q}_x,\hat{q}_y] = -ieB_z/\hbar$, Eq.~(\ref{effH}) can be readily derived from the three-fold rotational symmetry in TMDs, 
which gives $\Pi^x_{nm} = \pm i \Pi^y_{nm}$ \cite{PRR2019}.

Substituting Eq.~(\ref{effH}) into Eq.~(\ref{Lutt}), we obtain the equation for a free electron in the presence of $\mathbf{B}=(0,0,B_z)$, 
with energy levels shifted by a term linear in the magnetic field, given by $-\mathbf{m}_{n\mathbf{K}} \cdot \mathbf{B}$. 
Thus, solving Eq.~(\ref{Lutt}) and adding on $\mathbb{H}^{spin}$  gives LLs with energies
\begin{eqnarray}
\epsilon_N = E_{n\mathbf{K}}+(N+\frac{1}{2})\hbar\omega_c - g^{\text{orb}}_{n\mathbf{K}} \mu_B B_z + g_s s_z \mu_B B_z
\label{LLenergies}
\end{eqnarray}
where the cyclotron frequency $\omega_c=eB_z/m^*$, the quantum number $\alpha$ in Eq.~(\ref{Lutt}) is now the LL index $N=0,1,2,...$, 
and the single-band g-factor is defined as $g^{\text{orb}}_{n\mathbf{K}} \mu_B = m^z_{n\mathbf{K}}$. 
Notably, our result shows that the LL indices are symmetric in $K$ and $K'$, different from the results obtained from the
massive Dirac fermion model where the LL indices are valley-dependent \cite{LL2013,LLPRB2013}. The only valley-dependent term in Eq.~(\ref{LLenergies}) is $-g^{\text{orb}}_{n\mathbf{K}} \mu_B B_z$. 

We also highlight that the Bloch states are reorganized into LLs, so that the energies $\epsilon_N$ depend on the LL indices $N$, 
and not on the crystal momenta, which are not rigorously defined in the presence of an external magnetic field. 
The orbital magnetic moment as defined in Eq.~(\ref{defm}) should be evaluated exactly at the valley extrema. 
It determines the Zeeman shift for all the LLs (Eq.~(\ref{LLenergies})) that are located 
within the energy range in which the original Bloch state band structure is quadratic (Eq.~(\ref{k-dot-p})).

In contrast to previous work, we do not assume any specific form of the Hamiltonian.
Our derivation using the LK approximation results directly in an energy expression that includes both the LLs and Zeeman effects. 
Compared with the approach by Korm\'{a}nyos et al., \cite{NewJPhys} we start from a general Hamiltonian without the need to use a particular multiband $\mathbf{k} \cdot \mathbf{p}$ model.
Thus, the atomic contribution to the magnetic moment in the Zeeman effect is included automatically (see Section IIB for more details).
Our final expression for LLs in Eq.~(\ref{LLenergies}) can be directly used to interpret relevant experiments after obtaining the effective mass and $g^{\text{orb}}_{n\mathbf{K}}$ from first principles.
The limitation of this approach is that only terms up to second order in $\mathbf{q}$ are included in the expression for energy bands in Eq.~(\ref{k-dot-p}). 
So Eq.~(\ref{LLenergies}) is only a solution in the parabolic region of the energy extrema.
Applying the LK approximation to the full band would lead to the Hofstadter butterfly spectrum \cite{Hof}.

We now discuss another form of Eq.~(\ref{defm}). Using the result from perturbation theory that
\begin{eqnarray}
\partial_{\mathbf{k}} u_{n\mathbf{k}} = \frac{\hbar}{m_e}\sum_{m} \frac{<u_{m\mathbf{k}}|\bm{\pi}|u_{n\mathbf{k}}>}{E_{n\mathbf{k}}-E_{m\mathbf{k}}} u_{m\mathbf{k}}
\label{pupk}
\end{eqnarray}
$\mathbf{m}_{n\mathbf{K}}$ in Eq.~(\ref{defm}) can be written as
\begin{eqnarray}
\mathbf{m}_{n\mathbf{K}}=-\frac{ie}{2\hbar}<\partial_{\mathbf{k}} u_{n\mathbf{k}}|\times [H_{\mathbf{k}} - E_{n\mathbf{k}}]|\partial_{\mathbf{k}} u_{n\mathbf{k}}>|_{\mathbf{k}=\mathbf{K}}
\label{semim}
\end{eqnarray}
which has the same form as the semiclassical formula for $\mathbf{m}_{n\mathbf{K}}$ derived in Ref. \cite{Niu1996} in the absence of SOC.
We emphasize that Eq.~(\ref{semim}) is the orbital magnetic moment of single bands at non-degenerate band extrema with SOC, 
which is different from the total orbital magnetic moment that involves summation over all Bloch states \cite{VanPRB}.
Eq.~(\ref{semim}) is used instead of Eq.~(\ref{defm}) for our calculations of the g-factor, because the form of the orbital magnetic moment in Eq.~(\ref{semim}) 
holds also when $V(\mathbf{r})$ in Eq.~(\ref{HB}) is replaced by a non-local potential $V_{NL}(\mathbf{r},\mathbf{r}')$. 
This is in contrast to the expression in Eq.~(\ref{defm}), which requires the potential in the Hamiltonian to be local.

In the formalism presented above, we have used the property of locality in the periodic potential twice. 
Firstly, in the original proof of the LK approximation, a local periodic potential $V(\mathbf{r})$ is used in Eq.~(\ref{HB}). 
However, the LK approximation can be easily extended to the non-local case and one may refer to the discussion in Appendix \ref{AppB}.
Secondly, in the expansion (Eq.~(\ref{k-dot-p})) of energy bands in the absence of a magnetic field, 
the second order term (Eq.~(\ref{E2})) is valid only for a local potential. This means that Eq.~(\ref{defm}) is valid only for Hamiltonians with local potentials. 
For a non-local potential, we can express this second order term in the same form, but with $\mathbf{p}$ replaced by a more complicated expression (see Appendix \ref{AppC} for details).  
The resulting modified expression for Eq.~(\ref{defm}) is intractable for numerical implementation. 
Fortunately, the k-derivative of $u_{n\mathbf{k}}$ shown in Eq.~(\ref{pupk}) will also have $\mathbf{p}$ replaced by the same complicated expression, 
so that Eq.~(\ref{semim}) remains compact and unchanged.
Thus, Eq.~(\ref{semim}) should be used to compute the g-factors if one were to take into account the effects of non-local many-body corrections to the quasiparticle Hamiltonian. 

\subsection{Numerical Method}
In this subsection, we describe our numerical method to evaluate Eq.~(\ref{semim}) using first principles Hamiltonians taking into account exchange and correlation effects 
within DFT and within many-body perturbation theory using the GW approximation. 

Our numerical implementation for the DFT g-factors is performed in the  DFT code, QuantumESPRESSO \cite{QE}.
The $u_{n\mathbf{k}}$ in Eq.~(\ref{semim}) are obtained as the periodic part of the Bloch states, $\psi_{n\mathbf{k}}  = e^{i\mathbf{k}\cdot\mathbf{r}}u_{n\mathbf{k}}$,
which are the solutions of the Kohn-Sham equations, 
\begin{eqnarray}
H^{\text{DFT}}\psi_{n\mathbf{k}} = [\frac{\mathbf{p}^2}{2m_e}+V_{ion}+V_{H}+V_{xc}] \psi_{n\mathbf{k}}=E^{\text{DFT}}_{n\mathbf{k}}\psi_{n\mathbf{k}}
\end{eqnarray}
where $V_{ion}$, $V_{H}$ and $V_{xc}$ are the ionic, Hartree and exchange-correlation potentials.
The periodic potential $V$ in Eq.~(\ref{HB}) is taken to be the effective mean field potential $V^{MF}=V_{ion}+V_{H}+V_{xc}$ felt by an electron in the TMD material. 
For the DFT calculations in this work, the exchange-correlation functional is evaluated within the local density approximation (LDA) or the generalized gradient approximation (GGA) 
using the Perdew-Berke-Ernzerhof (PBE) \cite{PBE1996} parametrization. Any explicit dependence of the exchange-correlation functional on the current density \cite{PRL1987} is neglected. 
Similar approximations have been made for the computation of nuclear magnetic resonance chemical shifts, and good agreement with experiment was obtained \cite{NMR1996}. 
We use optimized norm-conserving pseudopotentials \cite{oncv} with an energy cutoff of $60$ Ry, and our ground state charge density is obtained using a $21\times21\times 1$ k-grid.

Then the key quantity $\partial_{\mathbf{k}} u_{n\mathbf{k}}$ is calculated as
\begin{eqnarray}
\partial_{\mathbf{k}} u_{n\mathbf{k}} = \frac{ e^{-i\theta} u_{n\mathbf{k}+d \mathbf{k}}- u_{n\mathbf{k}}}{d \mathbf{k}}
\label{p-transport}
\end{eqnarray}
where $e^{i\theta}= <u_{n\mathbf{k}}| u_{n\mathbf{k}+d \mathbf{k}}>/|<u_{n\mathbf{k}}| u_{n\mathbf{k}+d \mathbf{k}}>|$.
The term $ e^{i\theta}$ eliminates the random phase factor between $u_{n\mathbf{k}}$ and $u_{n\mathbf{k}+d\mathbf{k}}$, allowing us to use the parallel transport gauge.
Since the orbital magnetic moment and Berry curvature are both gauge-independent, such a parallel transport gauge will not affect our numerical results.
Thus, we directly evaluate Eq.~(\ref{semim}) using the DFT Kohn-Sham Hamiltonian, wave functions and eigenvalues.
We have checked the numerical convergence of both the Berry curvature and g-factor with respect to the magnitude of $d \mathbf{k}$,
which we have taken to be $10^{-5}$ times the reciprocal lattice constant.
Our self-consistent cycle is converged using an energy criterion of $10^{-8}$ Ry.
We comment that the above procedure remains the same whether or not SOC is included in the DFT Hamiltonian.

Although the Kohn-Sham wave functions are typically a good approximation to the quasiparticle wave functions, 
Kohn-Sham eigenvalues in general cannot be formally interpreted as quasiparticle energies. 
For example, DFT calculations using LDA and GGA exchange-correlation functionals usually underestimate the fundamental gap \cite{Louie1986,Louie2013PRL}.
Quasiparticle properties can be formulated rigorously within a Green's function approach, notably the GW approximation from many-body perturbation theory. 
So, it is important to evaluate the g-factors using the GW Hamiltonian, and we implement our numerical method in the BerkeleyGW package\cite{BGW}. 

DFT calculations are first performed to provide the mean field starting point for the GW calculation. 
The GW quasiparticle eigenvalues $E^{\text{QP}}_{n\mathbf{k}}$ are then obtained by solving the following equation:
\begin{eqnarray}
H^{\text{GW}}\psi_{n\mathbf{k}} = (H^{\text{DFT}}-V_{xc}+\Sigma)\psi_{n\mathbf{k}}=E^{\text{QP}}_{n\mathbf{k}}\psi_{n\mathbf{k}}
\end{eqnarray}
where $V_{xc}$ represents the exchange-correlation potential present in the DFT calculation, 
$\Sigma=iGW$ is the GW self-energy approximated by product of the Green's function and the screened Coulomb interaction $W(\mathbf{r},\mathbf{r}')$.
For the purpose of this paper, we ignore the energy dependence in $\Sigma$, using the so-called static Coulomb-hole-screened-exchange (COHSEX) approximation, 
introduced by Hedin \cite{Hedin1965,Louie1986}. In the COHSEX approximation, the self-energy consists of two terms: 
\begin{eqnarray}
\begin{split}
&\Sigma=\Sigma_{\text{SEX}}+\Sigma_{\text{COH}}\\
&\Sigma_{\text{SEX}}(\mathbf{r},\mathbf{r}') = -\sum_{n\mathbf{k}}^{occ} \psi_{n\mathbf{k}}(\mathbf{r}) \psi_{n\mathbf{k}}(\mathbf{r}')^* W(\mathbf{r},\mathbf{r}')\\
&\Sigma_{\text{COH}}(\mathbf{r},\mathbf{r}') =\frac{1}{2}\delta(\mathbf{r}-\mathbf{r}')[W(\mathbf{r},\mathbf{r}')-v(\mathbf{r},\mathbf{r}')] \
\end{split}
\end{eqnarray}
$\Sigma_{\text{SEX}}$ is the non-local screened-exchange interaction and $\Sigma_{\text{COH}}$ is the local Coulomb-hole term 
that represents the effect from the rearrangement of electrons around the quasiparticle \cite{Louie1986}. 
The dielectric matrix used to evaluate the screened Coulomb interaction is calculated within the random phase approximation.

Since the DFT wave functions are a good approximation to the quasiparticle wave functions\cite{Louie1986}, $\partial_{\mathbf{k}} u_{n\mathbf{k}}$ can be obtained from the DFT calculation. 
To evaluate the orbital magnetic moment, one simply needs to substitute the quasiparticle GW eigenvalues into $E_{n\mathbf{k}}$, and use $H^{\text{GW}}$ for the Hamiltonian $H$ in Eq.~(\ref{semim}):
\begin{eqnarray}
\mathbf{m}_{n\mathbf{K}}=-\frac{ie}{2\hbar}<\partial_{\mathbf{k}} u_{n\mathbf{k}}|\times [H^{\text{GW}}_{\mathbf{k}} - E^{\text{QP}}_{n\mathbf{k}}]|\partial_{\mathbf{k}} u_{n\mathbf{k}}>|_{\mathbf{k}=\mathbf{K}}
\end{eqnarray}
GW calculations with spinor wave functions are computationally challenging. Given that the electron spin is nearly $100\%$ polarized at each valley in the TMDs \cite{Qiu2016}, 
we can evaluate the GW correction to the g-factors using scalar wave functions, 
and then add these corrections to the g-factors computed using two-component spinor wavefunctions in DFT, thereby including SOC effects into GW g-factors. 
For all the GW calculations in this work, we use a cutoff of $35$ Ry for the dielectric matrix and a non-uniform sampling \cite{sub} of the Brillouin Zone starting with a $12 \times 12$ k-grid. 
Our g-factors are unchanged when the k-grid is increased to $18 \times 18$.

\section{Numerical Results and Discussion}
\subsection{DFT vs TB in ML MoS$_2$}

\begin{figure}
\centering
\includegraphics[width=7.5cm,clip=true]{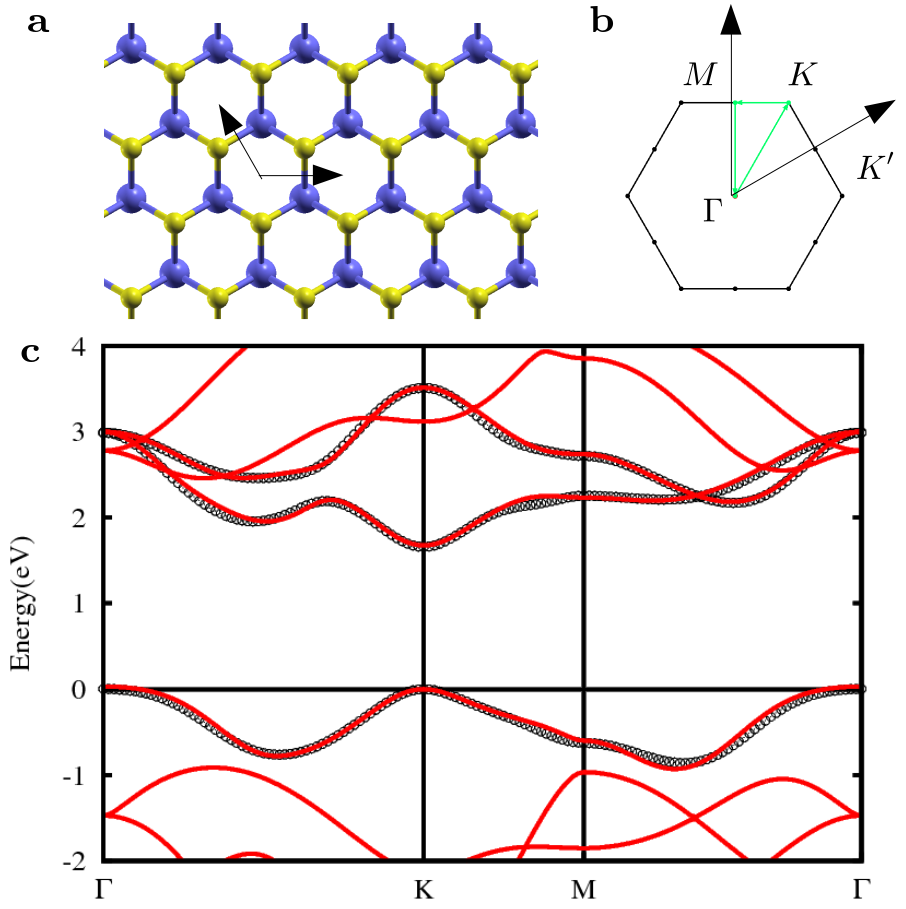}
\caption{(a) Top view of ML MoS$_2$. Blue and yellow balls represent Mo and S atoms, respectively, (b) Brillion zone with high symmetry k-path, 
(c) Band structure from DFT-PBE (red lines) and a three-band TB model\cite{TB2013} (black dots), without SOC.}
\label{figMo}
\end{figure}

\begin{table}
\caption{Single-band g-factors and Berry curvatures ($\Omega$ in \AA$^2$) at the $K$ point of ML MoS$_2$ computed using TB\cite{TB2013} and DFT methods.}
\begin{ruledtabular}
\begin{tabular}{l*{5}{c}}
          &TB &LDA &PBE &PBE-soc  \\
\hline
$g^{\text{orb}}_{c}$   &-3.98   &-1.94  &-1.91    & -2.05($\uparrow$) -1.83($\downarrow$) \\
$g^{\text{orb}}_{v}$   &-2.99   &-4.18  &-4.18    & -4.25($\uparrow$) -4.05($\downarrow$) \\
\hline
$\Omega_{c}$&-17.12&-9.01   &-8.82 &-9.97($\uparrow$) -8.00($\downarrow$) \\
$\Omega_{v}$& 15.82&9.95   & 9.72 &10.90($\uparrow$) 8.89($\downarrow$) \\
\end{tabular}
\label{tabTB}
\end{ruledtabular}
\end{table}

We compute the band structure, single band g-factors and Berry curvature in ML MoS$_2$, using DFT LDA and PBE calculations, as well as using a three-band TB model taken from Ref. \cite{TB2013}.
The atomic structure is shown in Figure~\ref{figMo}(a), where the lattice constant is taken to be the same as that used in Ref. \cite{TB2013}($3.19$ \AA).
The band structures calculated using DFT-PBE and using the TB model agree very well (Figure~\ref{figMo}(c)).
Table \ref{tabTB} shows our numerical results for the single band g-factors and Berry curvature at the $K$ valley. 
The TB results obtained in this work are consistent with those in Ref. \cite{TB2013}. 
Our DFT Berry curvature also agrees very well with DFT calculations in the literature, which were evaluated using the Kubo formula \cite{XD2012}. 
The good agreement of our results with the literature validate our numerical approach, and in particular, the use of Eq.~(\ref{p-transport}), which avoids the sum over unoccupied states.

From Table \ref{tabTB}, we find that the effect of SOC on the g-factors and Berry curvature is small, and roughly the same for the conduction and valence bands.
This is in contrast to the effect of SOC on the energy levels at the valleys, where the SOC splitting in the VBM is one order of magnitude larger than in the CBM\cite{Qiu2016}.
In the following section, we will see that the effect of SOC is larger for the g-factors in WSe$_2$. 
Next, for non-SOC calculations, one can see that both the g-factors and Berry curvature are almost the same for LDA and PBE, 
which suggests that at the DFT level, using local or semi-local exchange-correlation functionals makes no difference in predicting single band g-factors. 
However, the results computed using TB and DFT are quite different from one another.
Meanwhile, we notice that in Ref. \cite{TB11}, an eleven-band TB model gives $6.0$\AA$^2$ for $\Omega_{v}$, 
which is different from both the three-band TB model and DFT results.
These results strongly indicate that although the TB model can reproduce the first principles band structure, it is not reliable for computing the g-factors and Berry curvature.

To obtain physical insight into the origin of the orbital magnetic moment (g-factors) and the reason for the difference between first principles and TB results,
we consider a tight-binding basis $\beta_l = \frac{1}{\sqrt{N}}\sum_{\mathbf{R}} e^{i \mathbf{k} \cdot \mathbf{R}} \phi_l(\mathbf{r}-\mathbf{R})$ to expand the Bloch state,
\begin{eqnarray}
\psi_{n\mathbf{k}} = e^{i \mathbf{k} \cdot \mathbf{r}}u_{n\mathbf{k}} = \frac{1}{\sqrt{N}}\sum_{\mathbf{R}} e^{i \mathbf{k} \cdot \mathbf{R}} C_{n\mathbf{k}}^l \phi_l(\mathbf{r}-\mathbf{R})
\end{eqnarray}
where the Einstein summation convention is used. We then have,
\begin{eqnarray}
\begin{split}
\partial_{\mathbf{k}} u_{n\mathbf{k}} &= \frac{1}{\sqrt{N}}\sum_{\mathbf{R}} e^{i \mathbf{k} \cdot (\mathbf{R}-\mathbf{r})} \partial_{\mathbf{k}} C_{n\mathbf{k}}^l \phi_l(\mathbf{r}-\mathbf{R}) \\
&+ \frac{1}{\sqrt{N}}\sum_{\mathbf{R}} e^{i \mathbf{k} \cdot (\mathbf{R}-\mathbf{r})}i(\mathbf{R}-\mathbf{r}) C_{n\mathbf{k}}^l \phi_l(\mathbf{r}-\mathbf{R})
\end{split}
\label{partialu}
\end{eqnarray}
Substitution into Eq.~(\ref{semim}) gives three terms (Appendix \ref{AppD}), which we call valley (V), atomic (A) and cross (X) terms,
\begin{eqnarray}
\begin{split}
&\mathbf{m}_{n\mathbf{k}}^{(V)}=-\frac{ie}{2\hbar} [\partial_{\mathbf{k}} C_{n\mathbf{k}}^{l'}]^* \times [H_{l'l} - E_{n\mathbf{k}}S_{l'l}] [\partial_{\mathbf{k}} C_{n\mathbf{k}}^l]\\
&\mathbf{m}_{n\mathbf{k}}^{(A)}=\frac{e}{2m_e} [C_{n\mathbf{k}}^{l'}]^* \mathbf{L}_{l'l} C_{n\mathbf{k}}^l\\
&\mathbf{m}_{n\mathbf{k}}^{(X)}=-\frac{e}{m_e} Im \{ [\partial_{\mathbf{k}} C_{n\mathbf{k}}^{l'}]^* \times \bm{\Pi}_{l'l} [ C_{n\mathbf{k}}^l] \}
\end{split}
\label{partition}
\end{eqnarray}
The matrix elements are defined as $H_{l'l} = <\beta_{l'}|H|\beta_l>$, $S_{l'l} = <\beta_{l'}|\beta_l>$,
$\mathbf{L}_{l'l} = <\beta_{l'}|\mathbf{r}\times\bm{\pi}|\beta_l>$,
$\bm{\Pi}_{l'l} = <\beta_{l'}|\bm{\pi}|\beta_l>$. In the above, $\mathbf{k}$ is evaluated at $\mathbf{K}$ and $\mathbf{K}'=-\mathbf{K}$.
If the system has time reversal symmetry, each term fulfils $\mathbf{m}_{n\mathbf{K}}=-\mathbf{m}_{n\mathbf{K}'}$.
The expressions for valley and atomic terms depend, to a large extent, on the phase winding of the Bloch state and the parent atomic orbital angular momentum, respectively.
Thus, they can be regarded as analytical definitions for the valley and atomic terms used in the literature \cite{2d2019}.
The cross term, on the other hand, has not been discussed before.
(Note that the atomic term is also similar to the 'local' term in the total orbital magnetization discussed in Ref. \cite{solo2014}.) 
A similar result can be obtained for the Berry curvature \cite{JPCM2012}. There, the atomic term vanishes and we have,
\begin{eqnarray}
\begin{split}
&\Omega_{n\mathbf{k}}=i<\partial_{\mathbf{k}} u_{n\mathbf{k}}|\times|\partial_{\mathbf{k}} u_{n\mathbf{k}}>=\Omega_{n\mathbf{k}}^{V}+\Omega_{n\mathbf{k}}^{X}\\
&\Omega_{n\mathbf{k}}^{(V)} = i [\partial_{\mathbf{k}} C_{n\mathbf{k}}^{l'}]^* \times S_{l'l} [\partial_{\mathbf{k}} C_{n\mathbf{k}}^l]\\
&\Omega_{n\mathbf{k}}^{(X)} = 2 Re\{ { [\partial_{\mathbf{k}} C_{n\mathbf{k}}^{l'}]^* \times \mathbf{r}_{l'l} [ C_{n\mathbf{k}}^l] } \}
\end{split}
\label{partitionB}
\end{eqnarray}
where $\mathbf{r}_{l'l} = <\beta_{l'}|\mathbf{r}|\beta_l>$.

In standard TB or multi-band $\mathbf{k} \cdot \mathbf{p}$ Hamiltonian literature, $u_{n\mathbf{k}}$ is approximated as $ C_{n\mathbf{k}}^l $.
Thus, the second term in the expression for $\partial_{\mathbf{k}} u_{n\mathbf{k}}$ (Eq.~(\ref{partialu})) is ignored,
and only the valley term is present in the computed g-factors and Berry curvatures.
This explains why the atomic term is missing in some of the previous approaches \cite{LLPRB2013,NewJPhys,TB2013}.
In order to compute the two quantities using a TB model, one needs to evaluate all the terms in Eq.~(\ref{partition}) and Eq.~(\ref{partitionB}).
This is the major drawback of existing TB or $\mathbf{k} \cdot \mathbf{p}$ models for the description of the valley Zeeman effect in TMDs.

\subsection{DFT vs GW in WSe$_2$ and MoSe$_2$-WSe$_2$}

\begin{figure}
\centering
\includegraphics[width=7.5cm,clip=true]{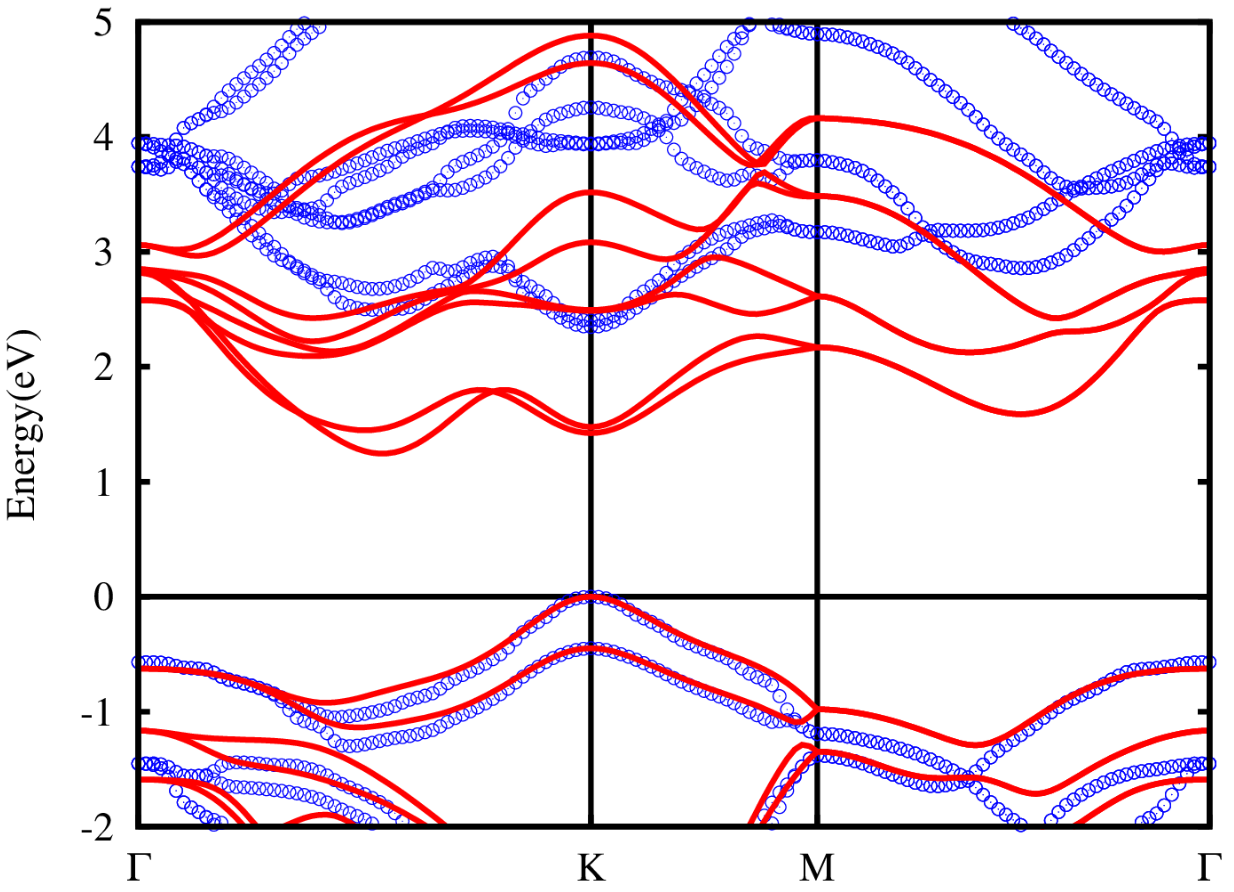}
\caption{PBE (red lines) and GW (blue dots) band structure of ML WSe$_2$, with SOC.}
\label{figW}
\end{figure}

\begin{table}
\caption{Single band and exciton g-factors of ML WSe$_2$ by DFT versus GW at $K$. $X0$ and $D0$ refer to the lowest-energy bright (spin-allowed) and dark (spin-forbidden) excitons, respectively. The numbers in brackets include the effects of a frequency-dependent BSE kernel.}
\begin{ruledtabular}
\begin{tabular}{l*{3}{c}}
$g^{\text{orb}}$         &PBE   &GW & Experiment \\
\hline
$g^{\text{orb}}_{c\uparrow}$         &-2.81    &-4.15  &        \\
$g^{\text{orb}}_{c\downarrow}$       &-1.90    &-3.24  &        \\
$g^{\text{orb}}_{v\uparrow}$         &-4.86    &-6.40  &       \\
$g^{\text{orb}}_{v\downarrow}$       &-4.17    &-5.71  &       \\
\hline
$g_{X0}$      &-4.10    &-4.50  & -3.7\cite{2d2015,NC2019} -4.3\cite{2d2019} -4.37\cite{NatPhys2015} -4.4\cite{Liu2019} \\
($\tilde{g}_{X0}$)          &         &(-4.26)&      \\
$g_{D0}$      &-9.92    &-10.32 & -9.3\cite{NC2019} -9.5\cite{Liu2019} -9.9\cite{PRR2019} \\
($\tilde{g}_{D0}$)          &         &(-9.76)                    \\
\end{tabular}
\label{tabML}
\end{ruledtabular}
\end{table}

Next, we perform DFT and GW calculations for single band and exciton g-factors. We choose ML WSe$_2$ as a prototypical example of 2H-phase TMDs.
Since there is no experimental data for the lattice constant in ML WSe$_2$, we use the experimental bulk lattice constant ($3.28$ \AA) for WSe$_2$.
Figure~\ref{figW} shows the PBE and GW band structures for ML WSe$_2$, with SOC effects. 
One can see that, consistent with GW calculations on similar systems \cite{Louie2013PRL}, 
the direct gap at $K$ is increased by $0.9$ eV due to the GW self-energy correction. Table~\ref{tabML} shows the g-factors from DFT and GW, including SOC effects.  
Compared to DFT-PBE, GW increases the magnitude of the single band g-factors significantly.
However, if one were to simply use Eq.~(\ref{defm}) for computing the orbital magnetic moment, 
one would immediately expect that the GW g-factors would be smaller in magnitude than the DFT-PBE values, because of the larger GW band gap (Figure~\ref{figW}).
As discussed in Section (IIA), the self-energy correction in the GW approximation is a non-local operator, 
so that one cannot simply use GW quasiparticle energies in Eq.~(\ref{defm}) to compute the g-factors.
The correct way is to use Eq.~(\ref{semim}), as described in Section (II), because the compact form of Eq.~(\ref{semim}) is unchanged in the presence of a non-local potential.
Noticing the fact that LDA and PBE give rather similar g-factors while GW changes the g-factors significantly, 
we expect that non-local exchange and correlation effects play an important role in the prediction of the single band g-factors.
Note that such effects cannot be included 
in g-factors in any TB or $\mathbf{k}\cdot\mathbf{p}$ model.  
Our implementation of Eq.(~\ref{semim}) paves the way for a deeper understanding of many-body effects on single band g-factors.

We also deduce the exciton g-factors to compare with experimental results. 
In the literature, experimentalists have measured the exciton Laud\'e g-factor, $g_{X0}$, 
defined as \cite{NatPhys2015,2d2015,2d2019},
\begin{eqnarray} 
\Delta E_{X0}= E_{X0}(\sigma+) - E_{X0}(\sigma-) = g_{X0}\mu_B B_z
\end{eqnarray}
where $\Delta E_{X0}$ is the Zeeman splitting of the first bright neutral exciton (X0) photoluminesence peak in H-phase TMDs in the presence of an out-of-plane magnetic field $B_z$.
According to our first principles calculations,
\begin{eqnarray}
<\psi_{c,\pm\mathbf{K}}|p_x|\psi_{v,\pm\mathbf{K}}>=\pm i <\psi_{c,\pm\mathbf{K}}|p_y|\psi_{v,\pm\mathbf{K}}>
\label{selML} 
\end{eqnarray}
in ML WSe$_2$. Thus, the conduction and valence bands in the $K$($K'$) valley couple to $\sigma+$($\sigma-$) circularly polarized light. 
From established excitonic physics \cite{Louie2013PRL}, we know that $X0$ is highly localised at the $K$ and $K'$ valleys, 
within the region in which the Bloch state band structure is quadratic, allowing us to use Eq.~(\ref{LLenergies}) for the energies of the states involved in the $X0$ exciton. 
Thus, using the fact that single-particle g-factors at $K$ and $K'$ have the same magnitude but opposite sign, 
one finds that the bright exciton g-factor $g_{X0} = 2(g^{\text{orb}}_{v\uparrow}-g^{\text{orb}}_{c\uparrow})$,
and the dark exciton g-factor $g_{D0} = 2(g^{\text{orb}}_{v\uparrow}-g^{\text{orb}}_{c\downarrow})-4$, where $-4$ comes from spin contributions.
Table~\ref{tabML} lists the bright and dark exciton g-factors derived using the spin-allowed and spin-forbidden transitions.
We obtain GW g-factors of $-4.50$ and $-10.32$ for the lowest energy spin-allowed and spin-forbidden transitions in ML WSe$_2$, respectively.
Compared to DFT, GW increases the absolute value of the single band g-factors, keeping the exciton g-factors almost unchanged.
One can see that the dark exciton g-factor is slightly larger than $g_{X0}-4$ due to the SOC effect on the single-band g-factors.
These exciton g-factors, as well as those obtained from DFT, are in good agreement with experiment (Table \ref{tabML}). 

The effects of electron-hole interactions on the exciton g-factors have been discussed by some authors \cite{Rubio}. 
Our consideration of electron-hole interactions will be published separately \cite{pre}. 
Briefly, we consider the effect of frequency-dependence in the kernel used in the Bethe-Salpeter equation (BSE), and, using a plasmon pole approximation, 
obtain a small reduction of the estimated g-factors by about $5.7\%$. This reduction gives exciton g-factors in slightly better agreement with experiment (Table~\ref{tabML}). 

\begin{figure*}
\centering
\includegraphics[width=15.0cm,clip=true]{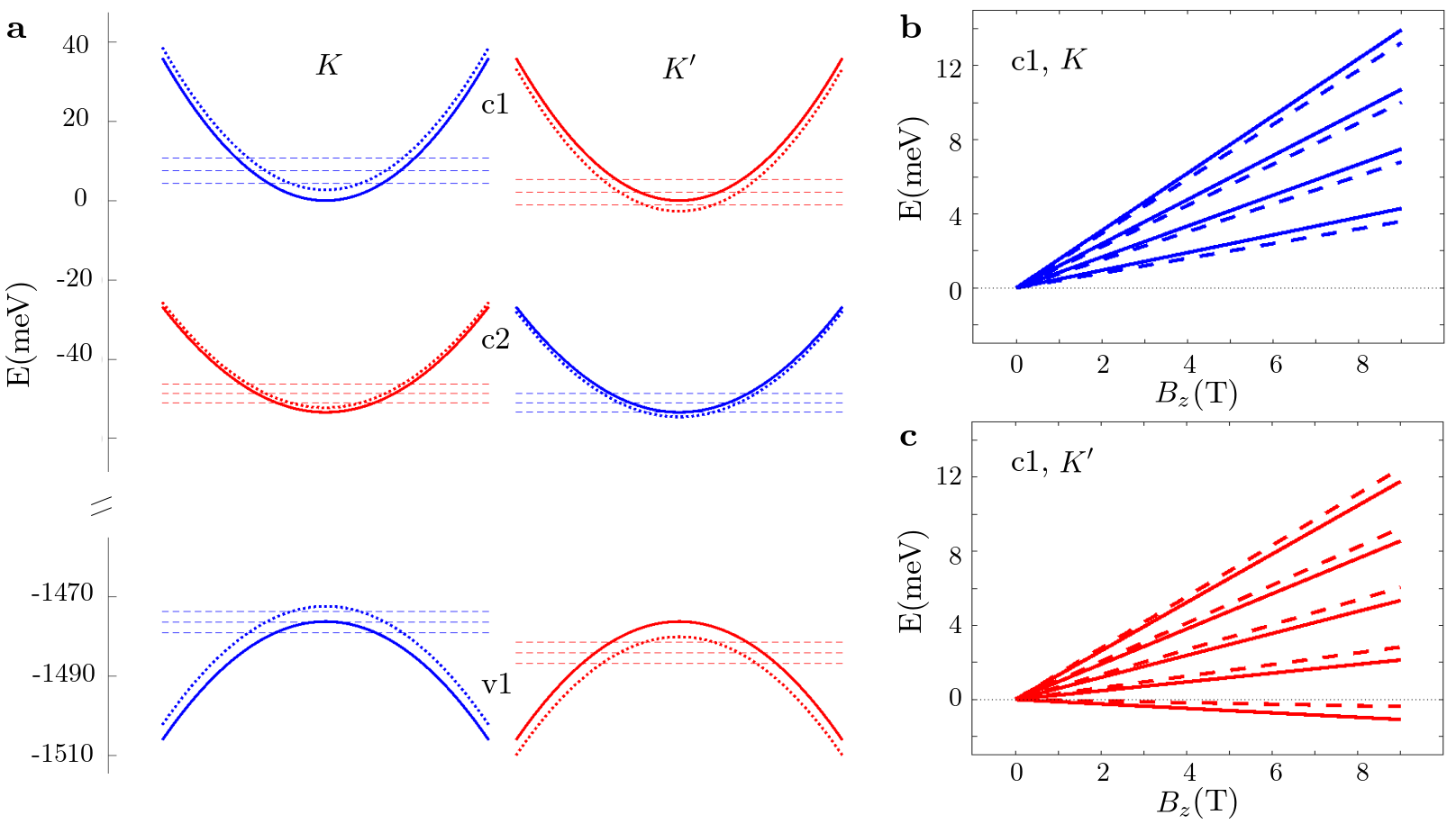}
\caption{Energy levels predicted at $K$ and $K'$ in ML WSe$_2$ in the presence of a uniform external magnetic field.
(a) Horizontal dashed lines denote the energy levels (LLs) obtained using a magnetic field of $9$ T.
Solid curves represent the energy bands in the absence of a magnetic field, and dashed curves represent the Zeeman-shifted bands (ignoring LL effects). GW g-factors are used for this plot; a plot using DFT g-factors looks similar.
(b-c) Energy of band $c1$ as a function of magnetic field strength at (b) $K$ and (c) $K'$. Solid lines and dashed lines are computed using GW and DFT-PBE g-factors, respectively. Blue denotes spin down and red denotes spin up.}
\label{figLL}
\end{figure*}

Using our computed effective masses of $0.32$, $0.44$ and $-0.39$ $m_e$ for the $c1$, $c2$ and $v1$ bands, respectively (see Figure~\ref{figLL}),
we predict from first principles the energy levels at the valleys including both LL and Zeeman effects (Eq.~(\ref{LLenergies})), and our results are shown in Figure~\ref{figLL}.
Both the LL diagram in Figure~\ref{figLL}(a) and the energies of $c1$ as a function of $B_z$ in Figure~\ref{figLL}(b-c) are in excellent agreement with
the experimental results in Ref. \cite{Mak} (the definitions of $K$ and $K'$ are reversed there).
In particular, our results are consistent with the conclusion in Ref. \cite{Mak} that the LLs are spin- and valley-polarized.
These energy levels include both the LL and Zeeman effects, computed entirely from first principles, and the g-factors include not only the spin and valley terms, 
but also the atomic and cross terms, in contrast to previous work\cite{LLPRB2013,NewJPhys}.

Note that the $N=0$ LL in Eq.~(\ref{LLenergies}) has a zero-point energy $\hbar\omega_c/2$, 
relative to the Bloch state band extrema predicted by taking into account the Zeeman terms only (see Figure~\ref{figLL}). 
If we were to assume that the exciton involves only the VBM and CBM wave functions at $K$ and $K'$, 
we would obtain an additional shift in the exciton (photoluminescence) energies due to the zero-point energies, 
which shift the VBM and CBM values to the LL energies. These terms are independent of the valley index. 
However, established excitonic physics shows that the $X0$ exciton involves transitions in a small region of the Brillouin Zone around $K$ \cite{Louie2013PRL}. 
The degeneracy of each LL in this region is the number of Bloch states in the original quadratic band dispersion within an energy range of $\pm\hbar\omega_c/2$ from this LL. 
Meanwhile, the allowed optical transitions are between conduction and valence band LLs with the same index \cite{Roth1959}.
Thus, the energies of the states involved in the $X0$ transition are reorganized from the quadratic band dispersion to quantized LL energies, 
and on average, this reorganization does not shift the exciton energies. The shift in exciton energy at each valley results from the Zeeman terms only.
This is also what is observed in experiments \cite{2d2015,NatPhys2015,Xu2019} where the shift in exciton energy at $K$ is equal and opposite to that at $K'$.

\begin{figure}
\centering
\includegraphics[width=7.5cm,clip=true]{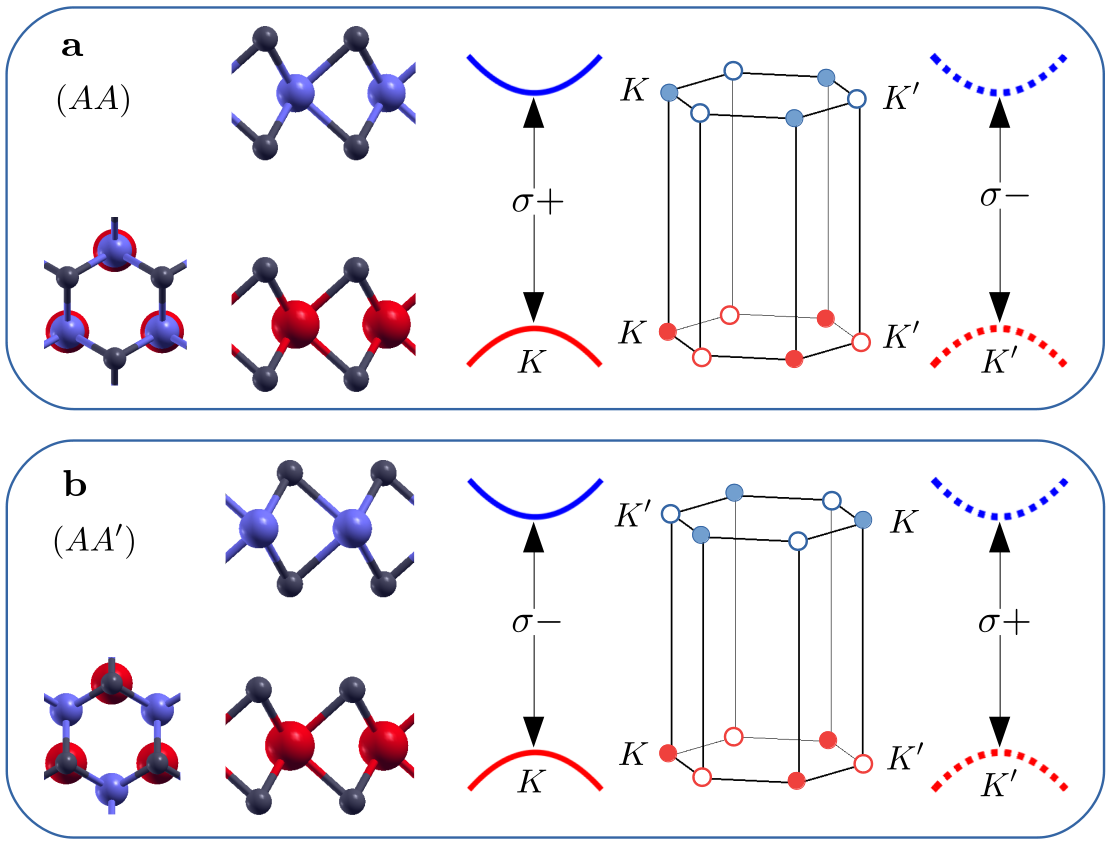}
\caption{Schematic figure for interlayer optical transitions in (a) $AA$ and (b) $AA'$ stacking orders in the MoSe$_2$-WSe$_2$ heterostructures. 
Blue and red balls represent Mo and W atoms, respectively.
Blue and red curves represent the conduction band in MoSe$_2$ layer and valence band in WSe$_2$ layer, respectively. 
Solid and dashed lines represent spin up and spin down states, respectively. The spin-allowed transitions in each valley are labeled by $\sigma+$ or $\sigma-$, indicative of the valley-dependent optical selection rules in these systems.}
\label{figMW}
\end{figure}

\begin{table}
\caption{Single band and interlayer spin-allowed exciton g-factors in $AA$-stacked and $AA'$-stacked MoSe$_2$-WSe$_2$ bilayers. 
Atomic symbols in brackets indicate which layer the corresponding band belongs to.}
\begin{ruledtabular}
\begin{tabular}{l*{4}{c}}
&$g^{\text{orb}}$         &PBE   &GW & Experiment \\
\hline
&$g^{\text{orb}}_{c\uparrow}$(Mo)     &-1.75    &-2.63  &        \\
$AA$&$g^{\text{orb}}_{v\uparrow}$(W)     &-4.84    &-6.14  &       \\
&$g_{X0}$                     &-6.18    &-7.02  & 6.72\cite{Xu2019} \\
\hline
&$g^{\text{orb}}_{c\uparrow}$(Mo)     &1.56    &2.43  &        \\
$AA'$&$g^{\text{orb}}_{v\uparrow}$(W)     &-4.69    &-5.81  &       \\
&$g_{X0}$                    &12.50    &16.48  & -15.79\cite{Xu2019} -15.1\cite{NC2017} \\
\end{tabular}
\label{tabAA}
\end{ruledtabular}
\end{table}

We also compute the g-factor for excitons corresponding to the lowest energy spin-allowed transitions in MoSe$_2$-WSe$_2$ heterobilayers in both $AA$ and $AA'$ stacking configurations, 
using the lattice constant in the ML WSe$_2$ calculations. These correspond, respectively, to twist angles of close to 60$^\circ$ and 0$^\circ$ in Ref. \cite{Xu2019} (see Figure~\ref{figMW}). 
Figure~\ref{figMW} shows a schematic of the interlayer transitions in MoSe$_2$-WSe$_2$ heterobilayers.
For the $AA$-stacked heterobilayer, the $K$ valley of the WSe$_2$ layer is aligned with the $K$ valley of the MoSe$_2$ layer, and similarly for the $K'$ valleys.
Thus, the valley optical selection rule is the same as that in the MLs (Eq.~(\ref{selML})).  
However, for the $AA'$-stacked heterobilayer, the $K$($K'$) valley of WSe$_2$ layer is aligned with $K'$($K$) valley of MoSe$_2$.
Defining the $K$ valley of the heterostructure as that for the WSe$_2$ layer, we obtain the following selection rule: 
\begin{eqnarray}
<\psi_{c,\pm\mathbf{K}}|p_x|\psi_{v,\pm\mathbf{K}}>=\mp i <\psi_{c,\pm\mathbf{K}}|p_y|\psi_{v,\pm\mathbf{K}}> 
\end{eqnarray}
and as a result, the interlayer exciton g factor for the $AA'$-stacked heterobilayer is $g_{X0} = 2(g^{\text{orb}}_{c\uparrow}-g^{\text{orb}}_{v\uparrow})$.

Table~\ref{tabAA} shows the relavent single-band g-factors, and the deduced exciton g-factors for MoSe$_2$-WSe$_2$ heterobilayers.
It is clear that GW gives interlayer exciton g-factors in better quantitative agreement with experiment than DFT. 
This better agreement is related to the larger magnitude of the single-band g-factors predicted by GW, 
underscoring the importance of including many-body effects in the quantitative understanding of the single-band g-factors and exciton g-factors.  
We note that interlayer hybridization should be explicitly taken into account to obtain quantitatively accurate exciton g-factors. 
If the single-band g-factors from isolated MoSe$_2$ and WSe$_2$ MLs are used to deduce the g-factors in the heterobilayer, 
the GW exciton g-factors will be $-7.14$ and $17.92$ for $AA$ and $AA'$-stacked systems. These are larger in magnitude than the values calculated directly from the heterostructure.

It is interesting to note that although the definition of the $K$ or $K'$ valley is arbitrary, 
the sign of the final exciton g-factor is determined once the sign convention of the external magnetic field and the light polarization are fixed.
The signs of the exciton g-factors we computed for the $AA$ and $AA'$ stackings are opposite to those in experiment\cite{Xu2019}, where only the twist angles were described. 
In the preparation of this manuscript, we became aware of Ref. \cite{arXiv2020} where the authors show that 
shifting the MoSe$_2$ layer to align its Se atom with the hollow site of the WSe$_2$ layer would change the sign of interlayer exciton g-factors.
We comment that this is related to differences in the valley selection rules between the two sets of structures, 
which originate from the differences in three-fold rotation centers of the two layers \cite{PRR2019}.

\section{Conclusions and outlook}
In conclusion, we have used the Luttinger-Kohn (LK) approximation to provide a unified description of the Zeeman effect and LLs in 2D TMDs, 
with both effects being treated on an equal footing within the same general Hamiltonian. 
We extend the original LK approximation to treat a Hamiltonian with a non-local periodic potential, allowing us to take into account non-local exchange and correlation effects on the single band g-factors.
The resulting energy levels are Landau levels (LLs) with LL indices that are symmetric in the $K$ and $K'$ valleys. These LLs are shifted by a valley-dependent Zeeman term. 
We develop a numerical approach to compute the Berry curvature and single-particle g-factors at the band extrema for a general Hamiltonian, without the need for a sum over unoccupied states. 
Tight-binding (TB), DFT LDA/PBE, and static GW (COHSEX) Hamiltonians are used in our calculations, 
in order to illustrate the effect of using increasingly better and more sophisticated approximations to the quasiparticle Hamiltonian. 
The TB Berry curvatures and single-particle g-factors are very different from the DFT results. 
This is because the TB results include only the so-called valley terms. 
On the other hand, the inclusion of many-body non-local exchange and correlation effects within the GW approximation increases the magnitude of the single band g-factors significantly compared to DFT. 
Spin-orbit coupling is included perturbatively.
The resulting LL diagram and exciton g-factors predicted by the GW calculations agree well with experiment, for both ML TMDs as well as twisted heterobilayers. 
The LLs we predict are spin- and valley-polarized. 

An interesting open question is whether it is possible to include the energy dependence of the self-energy in the computation of the g-factors, using our current theoretical framework. 
In this manuscript, we have limited our considerations to the static GW approximation, with every equation used in our calculations rigorously derived.
However, one can consider extending the LK approximation to treat an energy-dependent Hamiltonian, and include dynamical effects in the GW self-energies. 
In particular, the fact that our evaluation of the single-band g-factor does not require a sum over unoccupied states makes such an effort computationally efficient.
The approach developed here sets the stage for the treatment of non-local energy-dependent self-energy effects on the single band g-factors and LL energies in TMDs.

\textit{Note added.} After this manuscript was submitted, a paper on first principles calculations of exciton g-factors in ML TMDs appeared in the arXiv \cite{arXivBSE}.
 
\section{Acknowledgments}
We gratefully acknowledge support from the National Research Foundation, Singapore, for funding under the NRF Fellowship (NRF-NRFF2013-07) 
and under the NRF medium-sized centre programme. Calculations were performed on the computational cluster in the Centre for Advanced 2D Materials and the National Supercomputing Centre, Singapore. \\

\appendix
\section{\label{AppA}Second order expansion of energy band for a local periodic potential}
In this appendix, we expand the energy band at a band extremum point $\mathbf{k}_0$. 
The primary purpose of this Appendix is to lay the groundwork for the discussion in Appendix ~\ref{AppC} and SOC is omitted for simplicity. 
Since the periodic potential is local, the single particle Hamiltonian $H_{\mathbf{k}}=e^{-i\mathbf{k}\cdot\mathbf{r}}He^{i\mathbf{k}\cdot\mathbf{r}}$ is, 
\begin{eqnarray}
H_{\mathbf{k}}=\frac{\mathbf{p}^2}{2m_e}+V
+\frac{\hbar}{m_e}\mathbf{k}\cdot\mathbf{p}+\frac{\hbar^2}{2m_e}\mathbf{k}^2
\label{locHk}
\end{eqnarray}
where $H_{\mathbf{k}} u_{n\mathbf{k}}= E_{n\mathbf{k}} u_{n\mathbf{k}}$.
In order to find the second order expansion of energy bands about $\mathbf{k}_0$, one needs to evaluate:
\begin{eqnarray}
H_{\mathbf{k}_0+\mathbf{q}}=H_{\mathbf{k}_0}+\frac{\hbar}{m_e}(\hbar\mathbf{k}_0+\mathbf{p})\cdot\mathbf{q}+\frac{\hbar^2}{2m_e}\mathbf{q}^2
\end{eqnarray}
where $\mathbf{q}$ is a small shift relative to $\mathbf{k}_0$. Since $\mathbf{k}_0$ is a band extremum, the energy expansion up to second order is:
\begin{eqnarray}
E_{n\mathbf{k}_0+\mathbf{q}} = E_{n\mathbf{k}_0} +\frac{\hbar^2}{2m_e}\mathbf{q}^2 + E_{n\mathbf{k}_0}^{(2)} (\mathbf{q})
\label{secondorder}
\end{eqnarray}
where
\begin{eqnarray}
E_{n\mathbf{k}_0}^{(2)} (\mathbf{q}) = (\frac{\hbar}{m_e})^2\sum_{m \neq n}\frac{|\mathbf{q}\cdot <u_{n\mathbf{k}_0}|\mathbf{p}|u_{m\mathbf{k}_0}>|^2}{E_{n\mathbf{k}_0}-E_{m\mathbf{k}_0}}
\end{eqnarray}
The coefficient of the linear term in $\mathbf{k}$ in Eq.~\ref{locHk} determines the matrix elements in the expression for $E_{n\mathbf{k}_0}^{(2)} (\mathbf{q})$, 
and is what determines the final expression for the orbital magnetic moment in Eq.(~\ref{defm}).

\section{\label{AppB}Luttinger-Kohn approximation for non-local periodic potential}
In the original proof of the Luttinger-Kohn approximation \cite{Lut1951}, 
the wave function in the presence of an external magnetic field $\mathbf{B}=\nabla \times \mathbf{A}$ 
is represented in a basis of Wannier functions $W_{n\mathbf{R}}(\mathbf{r})$ (in this Appendix, we adopt the system of units used in Ref.\cite{Lut1951}):
\begin{eqnarray}
\psi_{\alpha}(\mathbf{r})=\sum_{\mathbf{R}}\Psi_{\alpha}(\mathbf{R})e^{ieG(\mathbf{R},\mathbf{r})/c\hbar}W_{n\mathbf{R}}(\mathbf{r}).
\end{eqnarray}
where
\begin{eqnarray}
G(\mathbf{R},\mathbf{r}) = \int_{\mathbf{R}}^{\mathbf{r}} \mathbf{A}(\mathbf{\eta}) d\mathbf{\eta}
\end{eqnarray}
The proof involves the evaluation of $\mathbb{H} \psi_{\alpha}(\mathbf{r})$, where $\mathbb{H} = \frac{(\mathbf{p}-e\mathbf{A}/c)^2}{2m_e}+V$
\begin{eqnarray}
\begin{split}
\mathbb{H} \psi_{\alpha}(\mathbf{r})&=\sum_{\mathbf{R}} \Psi_{\alpha}(\mathbf{R}) [\frac{(\mathbf{p}-e\mathbf{A}/c)^2}{2m_e}+V] e^{ie G(\mathbf{R},\mathbf{r})/c\hbar} W_{n\mathbf{R}}(\mathbf{r})\\
&=\sum_{\mathbf{R}} \Psi_{\alpha}(\mathbf{R}) e^{ie G(\mathbf{R},\mathbf{r})/c\hbar} [\frac{\mathbf{p}^2}{2m_e}+V] W_{n\mathbf{R}}(\mathbf{r})
\end{split}
\end{eqnarray}
In the above equation, the central idea is to move $e^{ie G /c\hbar}$ from the right hand side of $\mathbb{H}$ to its left hand side,
where the author used the argument that the Wannier function is highly localised at $\mathbf{R}$.
When one has a non-local periodic potential $V_{NL}(\mathbf{r},\mathbf{r}')$, this same argument that the Wannier function is localised can be used to move $e^{ie G/c\hbar}$ also:
\begin{eqnarray}
[V_{NL}(\mathbf{r},\mathbf{r}'),e^{ie G(\mathbf{R},\mathbf{r})/c\hbar}]W_{n\mathbf{R}}(\mathbf{r}) \approx 0
\end{eqnarray}
so that
\begin{eqnarray}
\begin{split}
&(\mathbb{H}+V_{NL}) \psi_{\alpha}(\mathbf{r})=\\
&\sum_{\mathbf{R}} \Psi_{\alpha}(\mathbf{R}) e^{ie G(\mathbf{R},\mathbf{r})/c\hbar}[\frac{\mathbf{p}^2}{2m_e}+V+V_{NL}] W_{n\mathbf{R}}(\mathbf{r})
\end{split}
\end{eqnarray}
The rest of the proof is unchanged, and in this way, the Luttinger-Kohn approximation can be easily extended to treat a non-local periodic potential.

\section{\label{AppC}Second order expansion of energy band for a non-local potential}
A general non-local potential $V_{NL}(\mathbf{r},\mathbf{r}')$ can be expanded as \cite{Mott}:
\begin{eqnarray}
V_{NL}(\mathbf{r},\mathbf{r}') = U_0(\mathbf{r}) + U_1(\mathbf{r}) \mathbf{r}\cdot\mathbf{p}+ U_2(\mathbf{r}) (\mathbf{r}\cdot\mathbf{p})^2+...
\end{eqnarray}
where $U_i$ are all local components. Since $V_{NL}(\mathbf{r},\mathbf{r}')$ does not commute with $e^{i\mathbf{k}\cdot\mathbf{r}}$, 
there will be an additional term in $H_{\mathbf{k}}$ when there is a non-local potential. This term is given by
\begin{eqnarray}
e^{-i\mathbf{k}\cdot\mathbf{r}}V_{NL}(\mathbf{r},\mathbf{r}')e^{i\mathbf{k}\cdot\mathbf{r}}
=U_0(\mathbf{r}) + U_1(\mathbf{r}) \mathbf{r}\cdot(\mathbf{p}+\hbar\mathbf{k})+ U_2(\mathbf{r}) (\mathbf{r}\cdot(\mathbf{p}+\hbar\mathbf{k}))^2+...
\label{V_NLk}
\end{eqnarray}
Recall from Appendix ~\ref{AppA} that the term linear in $\mathbf{k}$ in Eq.~\ref{locHk} determines the matrix elements in the expression for $E_{n\mathbf{k}_0}^{(2)} (\mathbf{q})$. 
In Eq.~\ref{V_NLk}, there is also a term linear in $\mathbf{k}$, which we can simply write as $\mathbf{k}\cdot\mathbf{c}$.
Since all the algebra to derive Eq.(~\ref{semim}) in Section (IIA) does not depend on the form of $\mathbf{c}$, 
one can simply replace $\mathbf{p}$ by $\mathbf{c}$, just as we replaced $\mathbf{p}$ with $\bm{\pi}$ to include the SOC effect.
Therefore, we show that the form of Eq.(~\ref{semim}) remains the same when there is a non-local potential.
However, the momentum matrix in Eq.(~\ref{defm}) should be replaced by $\mathbf{c}$, for which no closed form is known.

\section{\label{AppD}Partition of magnetic moment}
We provide below a detailed derivation for the partition of the orbital magnetic moment. As in the main text, expanding the Bloch state using a tight-binding basis we have,
$\partial_{\mathbf{k}} u_{n\mathbf{k}} = \mathbf{A}+\mathbf{B}$
\begin{eqnarray}
\mathbf{A} = \frac{1}{\sqrt{N}}\sum_{\mathbf{R}} e^{i \mathbf{k} \cdot (\mathbf{R}-\mathbf{r})}i(\mathbf{R}-\mathbf{r}) C_{n\mathbf{k}}^l \phi_l(\mathbf{r}-\mathbf{R})
\label{termA}
\end{eqnarray}
\begin{eqnarray}
\mathbf{B} = \frac{1}{\sqrt{N}}\sum_{\mathbf{R}} e^{i \mathbf{k} \cdot (\mathbf{R}-\mathbf{r})} \partial_{\mathbf{k}} C_{n\mathbf{k}}^l \phi_l(\mathbf{r}-\mathbf{R})
\label{termB}
\end{eqnarray}
Inserting $\partial_{\mathbf{k}} u_{n\mathbf{k}}$ into Eq.(~\ref{semim}) we obtain 
$\mathbf{m}_{n\mathbf{k}}=\mathbf{m}_{n\mathbf{k}}^{(V)} + \mathbf{m}_{n\mathbf{k}}^{(X)} +\mathbf{m}_{n\mathbf{k}}^{(A)}$.

First it is easy to derive the expression for the valley term:
\begin{eqnarray}
\begin{split}
\mathbf{m}_{n\mathbf{k}}^{(V)}&=-\frac{ie}{2\hbar}<\mathbf{B}|\times[H_{\mathbf{k}} - E_{n\mathbf{k}}]|\mathbf{B}> \\
&=-\frac{ie}{2\hbar}\frac{1}{N}\int d\mathbf{r} \sum_{\mathbf{R}'}[ \partial_{\mathbf{k}}C^{l'}_{n\mathbf{k}} \phi_{l'}]^* e^{-i \mathbf{k}(\mathbf{R}'-\mathbf{r})} \times [H_{\mathbf{k}} - E_{n\mathbf{k}}] 
\sum_{\mathbf{R}}[\partial_{\mathbf{k}}C^{l}_{n\mathbf{k}} \phi_l]e^{i\mathbf{k}(\mathbf{R}-\mathbf{r})}\\
&=-\frac{ie}{2\hbar}[ \partial_{\mathbf{k}}C^{l'}_{n\mathbf{k}}]^* \frac{1}{N}\int d\mathbf{r} \sum_{\mathbf{R}'} \phi_{l'}^*e^{-i \mathbf{k}\cdot\mathbf{R}'} \times [H - E_{n\mathbf{k}}] 
\sum_{\mathbf{R}}\phi_{l}e^{i\mathbf{k}\cdot\mathbf{R}}[\partial_{\mathbf{k}}C^{l}_{n\mathbf{k}}]\\
&=-\frac{ie}{2\hbar} [\partial_{\mathbf{k}} C_{n\mathbf{k}}^{l'}]^* \times [H_{l'l} - E_{n\mathbf{k}}S_{l'l}] [\partial_{\mathbf{k}} C_{n\mathbf{k}}^l]
\end{split}
\end{eqnarray}
where $H_{l'l} = <\beta_{l'}|\hat{H}|\beta_l>$ and $S_{l'l} = <\beta_{l'}|\beta_l>$. Next for the cross term:
\begin{eqnarray}
\begin{split}
\mathbf{m}_{n\mathbf{k}}^{(X)}&=\frac{e}{\hbar} Im {<\mathbf{B}|\times[H_{\mathbf{k}} - E_{n\mathbf{k}}]|\mathbf{A}>} \\
&=\frac{e}{\hbar} Im\{ \frac{1}{N}\int d\mathbf{r} \sum_{\mathbf{R}'}[ \partial_{\mathbf{k}}C^{l'}_{n\mathbf{k}} \phi_{l'}]^* e^{-i \mathbf{k}(\mathbf{R}'-\mathbf{r})}
\times [H_{\mathbf{k}} - E_{n\mathbf{k}}] \sum_{\mathbf{R}}[C^l_{n\mathbf{k}} \phi_l]e^{i\mathbf{k}(\mathbf{R}-\mathbf{r})}i(\mathbf{R}-\mathbf{r}) \}\\
&=-\frac{e}{m_e} Im { [\partial_{\mathbf{k}} C_{n\mathbf{k}}^{l'}]^* \times \bm{\Pi}_{l'l} [ C_{n\mathbf{k}}^l] }
\end{split}
\label{xt}
\end{eqnarray}
where $\bm{\Pi}_{l'l} = <\beta_{l'}|\bm{\pi}|\beta_l>$ and $\bm{\pi}=\mathbf{p}+\frac{\hbar}{4m_e c^2}(\bm{\sigma}\times \nabla V)$. One needs to show that the expression below vanishes:
\begin{eqnarray}
\begin{split}
\int d\mathbf{r} \sum_{\mathbf{R}'}[ \partial_{\mathbf{k}}C^{l'}_{n\mathbf{k}} \phi_{l'}]^* e^{-i \mathbf{k}(\mathbf{R}'-\mathbf{r})} \times [H_{\mathbf{k}} - E_{n\mathbf{k}}] 
\sum_{\mathbf{R}}[C^l_{n\mathbf{k}} \phi_l]e^{i\mathbf{k}(\mathbf{R}-\mathbf{r})} \mathbf{R} 
=\sum_{\mathbf{R}}w(\mathbf{R}) = 0
\end{split}
\end{eqnarray}
or show that $w(\mathbf{R}) = -w(-\mathbf{R})$ where,
\begin{eqnarray}
w(\mathbf{R}) = \int d\mathbf{r} \sum_{\mathbf{R}'}[ \partial_{\mathbf{k}}C^{l'}_{n\mathbf{k}} \phi_{l'}(\mathbf{r}-\mathbf{R}')]^* e^{-i \mathbf{k}(\mathbf{R}'-\mathbf{r})} \times [H_{\mathbf{k}} - E_{n\mathbf{k}}] 
[C^l_{n\mathbf{k}} \phi_l(\mathbf{r}-\mathbf{R})]e^{i\mathbf{k}(\mathbf{R}-\mathbf{r})} \mathbf{R}
\end{eqnarray}
So we prove as below
\begin{eqnarray}
\begin{split}
w(-\mathbf{R}) &= -\int d\mathbf{r} \sum_{\mathbf{R}'}[ \partial_{\mathbf{k}}C^{l'}_{n\mathbf{k}} \phi_{l'}]^* e^{-i \mathbf{k}(\mathbf{R}'-\mathbf{r})} 
\times [H_{\mathbf{k}} - E_{n\mathbf{k}}][C^l_{n\mathbf{k}} \phi_l(\mathbf{r}+\mathbf{R})]e^{i\mathbf{k}(-\mathbf{R}-\mathbf{r})} \mathbf{R}\\
&=-\int d\mathbf{r} \sum_{\mathbf{R}'+2\mathbf{R}}[ \partial_{\mathbf{k}}C^{l'}_{n\mathbf{k}} \phi_{l'}(\mathbf{r}+2\mathbf{R}-2\mathbf{R}-\mathbf{R}')]^* 
e^{-i \mathbf{k}(\mathbf{R}'+2\mathbf{R}-2\mathbf{R}-\mathbf{r})} \\
&\times [H_{\mathbf{k}} - E_{n\mathbf{k}}] [C^l_{n\mathbf{k}} \phi_l(\mathbf{r}+2\mathbf{R}-\mathbf{R})]e^{i\mathbf{k}(\mathbf{R}-2\mathbf{R}-\mathbf{r})} \mathbf{R}\\
&=-\int d(\mathbf{r}+2\mathbf{R}) \sum_{\mathbf{R}''}[ \partial_{\mathbf{k}}C^{l'}_{n\mathbf{k}} \phi_{l'}(\mathbf{r}+2\mathbf{R}-\mathbf{R}'')]^* e^{-i \mathbf{k}(\mathbf{R}''-2\mathbf{R}-\mathbf{r})} \\
&\times [H_{\mathbf{k}} - E_{n\mathbf{k}}] [C^l_{n\mathbf{k}} \phi_l(\mathbf{r}+2\mathbf{R}-\mathbf{R})]e^{i\mathbf{k}(\mathbf{R}-2\mathbf{R}-\mathbf{r})} \mathbf{R}\\
&=-\int d\mathbf{r}' \sum_{\mathbf{R}''}[ \partial_{\mathbf{k}}C^{l'}_{n\mathbf{k}} \phi_{l'}(\mathbf{r}'-\mathbf{R}'')]^* e^{-i \mathbf{k}(\mathbf{R}''-\mathbf{r}')} 
\times [H_{\mathbf{k}} - E_{n\mathbf{k}}] [C^l_{n\mathbf{k}} \phi_l(\mathbf{r}'-\mathbf{R})]e^{i\mathbf{k}(\mathbf{R}-\mathbf{r}')} \mathbf{R}\\
&=-w(\mathbf{R})
\end{split}
\end{eqnarray}
where we have replaced $\mathbf{R}''=\mathbf{R}'+2\mathbf{R}$ and $\mathbf{r}'=\mathbf{r}+2\mathbf{R}$.
Next, using $[H, \mathbf{r}] = -i\hbar\bm{\pi}/m_e$ and $(H_\mathbf{k} - E_{n\mathbf{k}}) \psi_{n\mathbf{k}}=0$, we arrive at Eq.~(\ref{xt}).

Finally for the atomic term, the $\mathbf{R}$ and $\mathbf{R}'$ in the expression below can be removed using a similar procedure as for the cross term, and the final expression is simple.
\begin{eqnarray}
\begin{split}
\mathbf{m}_{n\mathbf{k}}^{(A)}&=-\frac{ie}{2\hbar} <\mathbf{A}|\times[H_{\mathbf{k}} - E_{n\mathbf{k}}]|\mathbf{A}> \\
&=-\frac{ie}{2\hbar} \frac{1}{N}\int d\mathbf{r} \sum_{\mathbf{R}'}[C^{l'}_{n\mathbf{k}} \phi_{l'}]^*[-i(\mathbf{R}'-\mathbf{r})] e^{-i \mathbf{k}(\mathbf{R}'-\mathbf{r})}
\times [H_{\mathbf{k}} - E_{n\mathbf{k}}] \sum_{\mathbf{R}}[C^l_{n\mathbf{k}} \phi_l]e^{i\mathbf{k}(\mathbf{R}-\mathbf{r})}i(\mathbf{R}-\mathbf{r}) \\
&=\frac{e}{2m_e} [C_{n\mathbf{k}}^{l'}]^* \mathbf{L}_{l'l} C_{n\mathbf{k}}^l
\end{split}
\end{eqnarray}
where $\mathbf{L}_{l'l} = <\beta_{l'}|\mathbf{r}\times\bm{\pi}|\beta_l>$.


\bibliographystyle{apsrev4-1}
\bibliography{EndNotev2}

\begin{thebibliography}{56}%
\makeatletter
\providecommand \@ifxundefined [1]{%
 \@ifx{#1\undefined}
}%
\providecommand \@ifnum [1]{%
 \ifnum #1\expandafter \@firstoftwo
 \else \expandafter \@secondoftwo
 \fi
}%
\providecommand \@ifx [1]{%
 \ifx #1\expandafter \@firstoftwo
 \else \expandafter \@secondoftwo
 \fi
}%
\providecommand \natexlab [1]{#1}%
\providecommand \enquote  [1]{``#1''}%
\providecommand \bibnamefont  [1]{#1}%
\providecommand \bibfnamefont [1]{#1}%
\providecommand \citenamefont [1]{#1}%
\providecommand \href@noop [0]{\@secondoftwo}%
\providecommand \href [0]{\begingroup \@sanitize@url \@href}%
\providecommand \@href[1]{\@@startlink{#1}\@@href}%
\providecommand \@@href[1]{\endgroup#1\@@endlink}%
\providecommand \@sanitize@url [0]{\catcode `\\12\catcode `\$12\catcode
  `\&12\catcode `\#12\catcode `\^12\catcode `\_12\catcode `\%12\relax}%
\providecommand \@@startlink[1]{}%
\providecommand \@@endlink[0]{}%
\providecommand \url  [0]{\begingroup\@sanitize@url \@url }%
\providecommand \@url [1]{\endgroup\@href {#1}{\urlprefix }}%
\providecommand \urlprefix  [0]{URL }%
\providecommand \Eprint [0]{\href }%
\providecommand \doibase [0]{http://dx.doi.org/}%
\providecommand \selectlanguage [0]{\@gobble}%
\providecommand \bibinfo  [0]{\@secondoftwo}%
\providecommand \bibfield  [0]{\@secondoftwo}%
\providecommand \translation [1]{[#1]}%
\providecommand \BibitemOpen [0]{}%
\providecommand \bibitemStop [0]{}%
\providecommand \bibitemNoStop [0]{.\EOS\space}%
\providecommand \EOS [0]{\spacefactor3000\relax}%
\providecommand \BibitemShut  [1]{\csname bibitem#1\endcsname}%
\let\auto@bib@innerbib\@empty
\bibitem [{\citenamefont {Xiao}\ \emph {et~al.}(2012)\citenamefont {Xiao},
  \citenamefont {Liu}, \citenamefont {Feng}, \citenamefont {Xu},\ and\
  \citenamefont {Yao}}]{Xiao2012}%
  \BibitemOpen
  \bibfield  {author} {\bibinfo {author} {\bibfnamefont {D.}~\bibnamefont
  {Xiao}}, \bibinfo {author} {\bibfnamefont {G.-B.}\ \bibnamefont {Liu}},
  \bibinfo {author} {\bibfnamefont {W.}~\bibnamefont {Feng}}, \bibinfo {author}
  {\bibfnamefont {X.}~\bibnamefont {Xu}}, \ and\ \bibinfo {author}
  {\bibfnamefont {W.}~\bibnamefont {Yao}},\ }\href@noop {} {\bibfield
  {journal} {\bibinfo  {journal} {Phys. Rev. lett.}\ }\textbf {\bibinfo
  {volume} {108}},\ \bibinfo {pages} {196802} (\bibinfo {year}
  {2012})}\BibitemShut {NoStop}%
\bibitem [{\citenamefont {Koperski}\ \emph {et~al.}(2017)\citenamefont
  {Koperski}, \citenamefont {Molas}, \citenamefont {Aroro}, \citenamefont
  {Nogajewski}, \citenamefont {Slobodeniuk}, \citenamefont {Faugeras},\ and\
  \citenamefont {Potemski}}]{Nanoph}%
  \BibitemOpen
  \bibfield  {author} {\bibinfo {author} {\bibfnamefont {M.}~\bibnamefont
  {Koperski}}, \bibinfo {author} {\bibfnamefont {M.~R.}\ \bibnamefont {Molas}},
  \bibinfo {author} {\bibfnamefont {A.}~\bibnamefont {Aroro}}, \bibinfo
  {author} {\bibfnamefont {K.}~\bibnamefont {Nogajewski}}, \bibinfo {author}
  {\bibfnamefont {A.~O.}\ \bibnamefont {Slobodeniuk}}, \bibinfo {author}
  {\bibfnamefont {C.}~\bibnamefont {Faugeras}}, \ and\ \bibinfo {author}
  {\bibfnamefont {M.}~\bibnamefont {Potemski}},\ }\href@noop {} {\bibfield
  {journal} {\bibinfo  {journal} {Nanophotonics}\ }\textbf {\bibinfo {volume}
  {6}},\ \bibinfo {pages} {1289} (\bibinfo {year} {2017})}\BibitemShut
  {NoStop}%
\bibitem [{\citenamefont {Yu}\ \emph {et~al.}(2015)\citenamefont {Yu},
  \citenamefont {Cui}, \citenamefont {Xu},\ and\ \citenamefont
  {Yao}}]{Wang2015}%
  \BibitemOpen
  \bibfield  {author} {\bibinfo {author} {\bibfnamefont {H.}~\bibnamefont
  {Yu}}, \bibinfo {author} {\bibfnamefont {X.}~\bibnamefont {Cui}}, \bibinfo
  {author} {\bibfnamefont {X.}~\bibnamefont {Xu}}, \ and\ \bibinfo {author}
  {\bibfnamefont {W.}~\bibnamefont {Yao}},\ }\href@noop {} {\bibfield
  {journal} {\bibinfo  {journal} {Natl. Sci. Rev.}\ }\textbf {\bibinfo {volume}
  {2}},\ \bibinfo {pages} {57} (\bibinfo {year} {2015})}\BibitemShut {NoStop}%
\bibitem [{\citenamefont {Mueller}\ and\ \citenamefont
  {Malic}(2018)}]{NPJ2018}%
  \BibitemOpen
  \bibfield  {author} {\bibinfo {author} {\bibfnamefont {T.}~\bibnamefont
  {Mueller}}\ and\ \bibinfo {author} {\bibfnamefont {E.}~\bibnamefont
  {Malic}},\ }\href@noop {} {\bibfield  {journal} {\bibinfo  {journal} {NPJ 2D
  Mater. Appl.}\ }\textbf {\bibinfo {volume} {2}},\ \bibinfo {pages} {29}
  (\bibinfo {year} {2018})}\BibitemShut {NoStop}%
\bibitem [{\citenamefont {Wang}\ \emph {et~al.}(2018)\citenamefont {Wang},
  \citenamefont {Chernikov}, \citenamefont {Glazov}, \citenamefont {Heinz},
  \citenamefont {Marie}, \citenamefont {Amand},\ and\ \citenamefont
  {Urbaszek}}]{Wang2018}%
  \BibitemOpen
  \bibfield  {author} {\bibinfo {author} {\bibfnamefont {G.}~\bibnamefont
  {Wang}}, \bibinfo {author} {\bibfnamefont {A.}~\bibnamefont {Chernikov}},
  \bibinfo {author} {\bibfnamefont {M.~M.}\ \bibnamefont {Glazov}}, \bibinfo
  {author} {\bibfnamefont {T.~F.}\ \bibnamefont {Heinz}}, \bibinfo {author}
  {\bibfnamefont {X.}~\bibnamefont {Marie}}, \bibinfo {author} {\bibfnamefont
  {T.}~\bibnamefont {Amand}}, \ and\ \bibinfo {author} {\bibfnamefont
  {B.}~\bibnamefont {Urbaszek}},\ }\href@noop {} {\bibfield  {journal}
  {\bibinfo  {journal} {Rev. Mod. Phys.}\ }\textbf {\bibinfo {volume} {90}},\
  \bibinfo {pages} {021001} (\bibinfo {year} {2018})}\BibitemShut {NoStop}%
\bibitem [{\citenamefont {Schaibley}\ \emph {et~al.}(2016)\citenamefont
  {Schaibley}, \citenamefont {Yu}, \citenamefont {Clark}, \citenamefont
  {Rivera}, \citenamefont {Ross}, \citenamefont {Seyler}, \citenamefont {Yao},\
  and\ \citenamefont {Xu}}]{XX2016}%
  \BibitemOpen
  \bibfield  {author} {\bibinfo {author} {\bibfnamefont {J.}~\bibnamefont
  {Schaibley}}, \bibinfo {author} {\bibfnamefont {H.}~\bibnamefont {Yu}},
  \bibinfo {author} {\bibfnamefont {G.}~\bibnamefont {Clark}}, \bibinfo
  {author} {\bibfnamefont {P.}~\bibnamefont {Rivera}}, \bibinfo {author}
  {\bibfnamefont {J.~S.}\ \bibnamefont {Ross}}, \bibinfo {author}
  {\bibfnamefont {K.~L.}\ \bibnamefont {Seyler}}, \bibinfo {author}
  {\bibfnamefont {W.}~\bibnamefont {Yao}}, \ and\ \bibinfo {author}
  {\bibfnamefont {X.}~\bibnamefont {Xu}},\ }\href@noop {} {\bibfield  {journal}
  {\bibinfo  {journal} {Nat. Rev. Mater.}\ }\textbf {\bibinfo {volume} {1}},\
  \bibinfo {pages} {16055} (\bibinfo {year} {2016})}\BibitemShut {NoStop}%
\bibitem [{\citenamefont {Srivastava}\ \emph {et~al.}(2015)\citenamefont
  {Srivastava}, \citenamefont {Sidler}, \citenamefont {Allain}, \citenamefont
  {Lembke}, \citenamefont {Kis},\ and\ \citenamefont
  {Imamo\u{g}lu}}]{NatPhys2015}%
  \BibitemOpen
  \bibfield  {author} {\bibinfo {author} {\bibfnamefont {A.}~\bibnamefont
  {Srivastava}}, \bibinfo {author} {\bibfnamefont {M.}~\bibnamefont {Sidler}},
  \bibinfo {author} {\bibfnamefont {A.~V.}\ \bibnamefont {Allain}}, \bibinfo
  {author} {\bibfnamefont {D.~S.}\ \bibnamefont {Lembke}}, \bibinfo {author}
  {\bibfnamefont {A.}~\bibnamefont {Kis}}, \ and\ \bibinfo {author}
  {\bibfnamefont {A.}~\bibnamefont {Imamo\u{g}lu}},\ }\href@noop {} {\bibfield
  {journal} {\bibinfo  {journal} {Nat. Phys.}\ }\textbf {\bibinfo {volume}
  {11}},\ \bibinfo {pages} {141} (\bibinfo {year} {2015})}\BibitemShut
  {NoStop}%
\bibitem [{\citenamefont {Wang}\ \emph {et~al.}(2017)\citenamefont {Wang},
  \citenamefont {Shan},\ and\ \citenamefont {Mak}}]{Mak}%
  \BibitemOpen
  \bibfield  {author} {\bibinfo {author} {\bibfnamefont {Z.}~\bibnamefont
  {Wang}}, \bibinfo {author} {\bibfnamefont {J.}~\bibnamefont {Shan}}, \ and\
  \bibinfo {author} {\bibfnamefont {K.~F.}\ \bibnamefont {Mak}},\ }\href@noop
  {} {\bibfield  {journal} {\bibinfo  {journal} {Nat. Nanotechnol.}\ }\textbf
  {\bibinfo {volume} {12}},\ \bibinfo {pages} {144} (\bibinfo {year}
  {2017})}\BibitemShut {NoStop}%
\bibitem [{\citenamefont {Yao}\ \emph {et~al.}(2008)\citenamefont {Yao},
  \citenamefont {Xiao},\ and\ \citenamefont {Niu}}]{Niu2008}%
  \BibitemOpen
  \bibfield  {author} {\bibinfo {author} {\bibfnamefont {W.}~\bibnamefont
  {Yao}}, \bibinfo {author} {\bibfnamefont {D.}~\bibnamefont {Xiao}}, \ and\
  \bibinfo {author} {\bibfnamefont {Q.}~\bibnamefont {Niu}},\ }\href@noop {}
  {\bibfield  {journal} {\bibinfo  {journal} {Phys. Rev. B}\ }\textbf {\bibinfo
  {volume} {77}},\ \bibinfo {pages} {235406} (\bibinfo {year}
  {2008})}\BibitemShut {NoStop}%
\bibitem [{\citenamefont {Xiao}\ \emph {et~al.}(2010)\citenamefont {Xiao},
  \citenamefont {Chang},\ and\ \citenamefont {Niu}}]{RMP2010}%
  \BibitemOpen
  \bibfield  {author} {\bibinfo {author} {\bibfnamefont {D.}~\bibnamefont
  {Xiao}}, \bibinfo {author} {\bibfnamefont {M.-C.}\ \bibnamefont {Chang}}, \
  and\ \bibinfo {author} {\bibfnamefont {Q.}~\bibnamefont {Niu}},\ }\href@noop
  {} {\bibfield  {journal} {\bibinfo  {journal} {Rev. Mod. Phys.}\ }\textbf
  {\bibinfo {volume} {82}},\ \bibinfo {pages} {1959} (\bibinfo {year}
  {2010})}\BibitemShut {NoStop}%
\bibitem [{\citenamefont {Cao}\ \emph {et~al.}(2012)\citenamefont {Cao},
  \citenamefont {Wang}, \citenamefont {Han}, \citenamefont {Ye}, \citenamefont
  {Zhu}, \citenamefont {Shi}, \citenamefont {Niu}, \citenamefont {Tan},
  \citenamefont {Wang}, \citenamefont {Liu},\ and\ \citenamefont
  {Feng}}]{NC2012}%
  \BibitemOpen
  \bibfield  {author} {\bibinfo {author} {\bibfnamefont {T.}~\bibnamefont
  {Cao}}, \bibinfo {author} {\bibfnamefont {G.}~\bibnamefont {Wang}}, \bibinfo
  {author} {\bibfnamefont {W.}~\bibnamefont {Han}}, \bibinfo {author}
  {\bibfnamefont {H.}~\bibnamefont {Ye}}, \bibinfo {author} {\bibfnamefont
  {C.}~\bibnamefont {Zhu}}, \bibinfo {author} {\bibfnamefont {J.}~\bibnamefont
  {Shi}}, \bibinfo {author} {\bibfnamefont {Q.}~\bibnamefont {Niu}}, \bibinfo
  {author} {\bibfnamefont {P.}~\bibnamefont {Tan}}, \bibinfo {author}
  {\bibfnamefont {E.}~\bibnamefont {Wang}}, \bibinfo {author} {\bibfnamefont
  {B.}~\bibnamefont {Liu}}, \ and\ \bibinfo {author} {\bibfnamefont
  {J.}~\bibnamefont {Feng}},\ }\href@noop {} {\bibfield  {journal} {\bibinfo
  {journal} {Nat. Commun.}\ }\textbf {\bibinfo {volume} {3}},\ \bibinfo {pages}
  {887} (\bibinfo {year} {2012})}\BibitemShut {NoStop}%
\bibitem [{\citenamefont {Wang}\ \emph {et~al.}(2015)\citenamefont {Wang},
  \citenamefont {Bouet}, \citenamefont {Glazov}, \citenamefont {Amand},
  \citenamefont {Ivchenko}, \citenamefont {Palleau}, \citenamefont {Marie},\
  and\ \citenamefont {Urbaszek}}]{2d2015}%
  \BibitemOpen
  \bibfield  {author} {\bibinfo {author} {\bibfnamefont {G.}~\bibnamefont
  {Wang}}, \bibinfo {author} {\bibfnamefont {L.}~\bibnamefont {Bouet}},
  \bibinfo {author} {\bibfnamefont {M.~M.}\ \bibnamefont {Glazov}}, \bibinfo
  {author} {\bibfnamefont {T.}~\bibnamefont {Amand}}, \bibinfo {author}
  {\bibfnamefont {E.~L.}\ \bibnamefont {Ivchenko}}, \bibinfo {author}
  {\bibfnamefont {E.}~\bibnamefont {Palleau}}, \bibinfo {author} {\bibfnamefont
  {X.}~\bibnamefont {Marie}}, \ and\ \bibinfo {author} {\bibfnamefont
  {B.}~\bibnamefont {Urbaszek}},\ }\href@noop {} {\bibfield  {journal}
  {\bibinfo  {journal} {2D Mater.}\ }\textbf {\bibinfo {volume} {2}},\ \bibinfo
  {pages} {034002} (\bibinfo {year} {2015})}\BibitemShut {NoStop}%
\bibitem [{\citenamefont {Chang}\ and\ \citenamefont {Niu}(1996)}]{Niu1996}%
  \BibitemOpen
  \bibfield  {author} {\bibinfo {author} {\bibfnamefont {M.-C.}\ \bibnamefont
  {Chang}}\ and\ \bibinfo {author} {\bibfnamefont {Q.}~\bibnamefont {Niu}},\
  }\href@noop {} {\bibfield  {journal} {\bibinfo  {journal} {Phys. Rev. B}\
  }\textbf {\bibinfo {volume} {53}},\ \bibinfo {pages} {7010} (\bibinfo {year}
  {1996})}\BibitemShut {NoStop}%
\bibitem [{\citenamefont {Rose}\ \emph {et~al.}(2013)\citenamefont {Rose},
  \citenamefont {Goerbig},\ and\ \citenamefont {Pi\'{e}chon}}]{LL2013}%
  \BibitemOpen
  \bibfield  {author} {\bibinfo {author} {\bibfnamefont {F.}~\bibnamefont
  {Rose}}, \bibinfo {author} {\bibfnamefont {M.~O.}\ \bibnamefont {Goerbig}}, \
  and\ \bibinfo {author} {\bibfnamefont {F.}~\bibnamefont {Pi\'{e}chon}},\
  }\href@noop {} {\bibfield  {journal} {\bibinfo  {journal} {Phys. Rev. B}\
  }\textbf {\bibinfo {volume} {88}},\ \bibinfo {pages} {125438} (\bibinfo
  {year} {2013})}\BibitemShut {NoStop}%
\bibitem [{\citenamefont {Cai}\ \emph {et~al.}(2013)\citenamefont {Cai},
  \citenamefont {Yang}, \citenamefont {Li}, \citenamefont {Zhang},
  \citenamefont {Shui}, \citenamefont {Yao},\ and\ \citenamefont
  {Niu}}]{LLPRB2013}%
  \BibitemOpen
  \bibfield  {author} {\bibinfo {author} {\bibfnamefont {T.}~\bibnamefont
  {Cai}}, \bibinfo {author} {\bibfnamefont {S.~A.}\ \bibnamefont {Yang}},
  \bibinfo {author} {\bibfnamefont {X.}~\bibnamefont {Li}}, \bibinfo {author}
  {\bibfnamefont {F.}~\bibnamefont {Zhang}}, \bibinfo {author} {\bibfnamefont
  {J.}~\bibnamefont {Shui}}, \bibinfo {author} {\bibfnamefont {W.}~\bibnamefont
  {Yao}}, \ and\ \bibinfo {author} {\bibfnamefont {Q.}~\bibnamefont {Niu}},\
  }\href@noop {} {\bibfield  {journal} {\bibinfo  {journal} {Phys. Rev. B}\
  }\textbf {\bibinfo {volume} {88}},\ \bibinfo {pages} {115140} (\bibinfo
  {year} {2013})}\BibitemShut {NoStop}%
\bibitem [{\citenamefont {Li}\ \emph {et~al.}(2013)\citenamefont {Li},
  \citenamefont {Zhang},\ and\ \citenamefont {Niu}}]{LLPRL2013}%
  \BibitemOpen
  \bibfield  {author} {\bibinfo {author} {\bibfnamefont {X.}~\bibnamefont
  {Li}}, \bibinfo {author} {\bibfnamefont {F.}~\bibnamefont {Zhang}}, \ and\
  \bibinfo {author} {\bibfnamefont {Q.}~\bibnamefont {Niu}},\ }\href@noop {}
  {\bibfield  {journal} {\bibinfo  {journal} {Phys. Rev. Lett.}\ }\textbf
  {\bibinfo {volume} {110}},\ \bibinfo {pages} {066803} (\bibinfo {year}
  {2013})}\BibitemShut {NoStop}%
\bibitem [{\citenamefont {Fallahazad}\ \emph {et~al.}(2016)\citenamefont
  {Fallahazad}, \citenamefont {Movva}, \citenamefont {Kim}, \citenamefont
  {Larentis}, \citenamefont {Taniguchi}, \citenamefont {Watanabe},
  \citenamefont {Banerjee},\ and\ \citenamefont {Tutuc}}]{LL2016}%
  \BibitemOpen
  \bibfield  {author} {\bibinfo {author} {\bibfnamefont {B.}~\bibnamefont
  {Fallahazad}}, \bibinfo {author} {\bibfnamefont {H.~C.~P.}\ \bibnamefont
  {Movva}}, \bibinfo {author} {\bibfnamefont {K.}~\bibnamefont {Kim}}, \bibinfo
  {author} {\bibfnamefont {S.}~\bibnamefont {Larentis}}, \bibinfo {author}
  {\bibfnamefont {T.}~\bibnamefont {Taniguchi}}, \bibinfo {author}
  {\bibfnamefont {K.}~\bibnamefont {Watanabe}}, \bibinfo {author}
  {\bibfnamefont {S.~K.}\ \bibnamefont {Banerjee}}, \ and\ \bibinfo {author}
  {\bibfnamefont {E.}~\bibnamefont {Tutuc}},\ }\href@noop {} {\bibfield
  {journal} {\bibinfo  {journal} {Phys. Rev. Lett.}\ }\textbf {\bibinfo
  {volume} {116}},\ \bibinfo {pages} {086601} (\bibinfo {year}
  {2016})}\BibitemShut {NoStop}%
\bibitem [{\citenamefont {Movva}\ \emph {et~al.}(2017)\citenamefont {Movva},
  \citenamefont {Fallahazad}, \citenamefont {Kim}, \citenamefont {Larentis},
  \citenamefont {Taniguchi}, \citenamefont {Watanabe}, \citenamefont
  {Banerjee},\ and\ \citenamefont {Tutuc}}]{LL2017}%
  \BibitemOpen
  \bibfield  {author} {\bibinfo {author} {\bibfnamefont {H.~C.~P.}\
  \bibnamefont {Movva}}, \bibinfo {author} {\bibfnamefont {B.}~\bibnamefont
  {Fallahazad}}, \bibinfo {author} {\bibfnamefont {K.}~\bibnamefont {Kim}},
  \bibinfo {author} {\bibfnamefont {S.}~\bibnamefont {Larentis}}, \bibinfo
  {author} {\bibfnamefont {T.}~\bibnamefont {Taniguchi}}, \bibinfo {author}
  {\bibfnamefont {K.}~\bibnamefont {Watanabe}}, \bibinfo {author}
  {\bibfnamefont {S.~K.}\ \bibnamefont {Banerjee}}, \ and\ \bibinfo {author}
  {\bibfnamefont {E.}~\bibnamefont {Tutuc}},\ }\href@noop {} {\bibfield
  {journal} {\bibinfo  {journal} {Phys. Rev. Lett.}\ }\textbf {\bibinfo
  {volume} {118}},\ \bibinfo {pages} {247701} (\bibinfo {year}
  {2017})}\BibitemShut {NoStop}%
\bibitem [{\citenamefont {Smole\'{n}ski}\ \emph {et~al.}(2019)\citenamefont
  {Smole\'{n}ski}, \citenamefont {Cotlet}, \citenamefont {Popert},
  \citenamefont {Back}, \citenamefont {Shimazaki}, \citenamefont {Kn\"{u}ppel},
  \citenamefont {Dietler}, \citenamefont {Taniguchi}, \citenamefont {Watanabe},
  \citenamefont {Kroner},\ and\ \citenamefont {Imamo\u{g}lu}}]{LL2019}%
  \BibitemOpen
  \bibfield  {author} {\bibinfo {author} {\bibfnamefont {T.}~\bibnamefont
  {Smole\'{n}ski}}, \bibinfo {author} {\bibfnamefont {O.}~\bibnamefont
  {Cotlet}}, \bibinfo {author} {\bibfnamefont {A.}~\bibnamefont {Popert}},
  \bibinfo {author} {\bibfnamefont {P.}~\bibnamefont {Back}}, \bibinfo {author}
  {\bibfnamefont {Y.}~\bibnamefont {Shimazaki}}, \bibinfo {author}
  {\bibfnamefont {P.}~\bibnamefont {Kn\"{u}ppel}}, \bibinfo {author}
  {\bibfnamefont {N.}~\bibnamefont {Dietler}}, \bibinfo {author} {\bibfnamefont
  {T.}~\bibnamefont {Taniguchi}}, \bibinfo {author} {\bibfnamefont
  {K.}~\bibnamefont {Watanabe}}, \bibinfo {author} {\bibfnamefont
  {M.}~\bibnamefont {Kroner}}, \ and\ \bibinfo {author} {\bibfnamefont
  {A.}~\bibnamefont {Imamo\u{g}lu}},\ }\href@noop {} {\bibfield  {journal}
  {\bibinfo  {journal} {Phys. Rev. Lett.}\ }\textbf {\bibinfo {volume} {123}},\
  \bibinfo {pages} {097403} (\bibinfo {year} {2019})}\BibitemShut {NoStop}%
\bibitem [{\citenamefont {Koperski}\ \emph {et~al.}(2019)\citenamefont
  {Koperski}, \citenamefont {Molas}, \citenamefont {Arora}, \citenamefont
  {Nogajewski}, \citenamefont {Bartos}, \citenamefont {Wyzula}, \citenamefont
  {Vaclavkova}, \citenamefont {Kossacki},\ and\ \citenamefont
  {Potemski}}]{2d2019}%
  \BibitemOpen
  \bibfield  {author} {\bibinfo {author} {\bibfnamefont {M.}~\bibnamefont
  {Koperski}}, \bibinfo {author} {\bibfnamefont {M.~R.}\ \bibnamefont {Molas}},
  \bibinfo {author} {\bibfnamefont {A.}~\bibnamefont {Arora}}, \bibinfo
  {author} {\bibfnamefont {K.}~\bibnamefont {Nogajewski}}, \bibinfo {author}
  {\bibfnamefont {M.}~\bibnamefont {Bartos}}, \bibinfo {author} {\bibfnamefont
  {J.}~\bibnamefont {Wyzula}}, \bibinfo {author} {\bibfnamefont
  {D.}~\bibnamefont {Vaclavkova}}, \bibinfo {author} {\bibfnamefont
  {P.}~\bibnamefont {Kossacki}}, \ and\ \bibinfo {author} {\bibfnamefont
  {M.}~\bibnamefont {Potemski}},\ }\href@noop {} {\bibfield  {journal}
  {\bibinfo  {journal} {2D Mater.}\ }\textbf {\bibinfo {volume} {6}},\ \bibinfo
  {pages} {015001} (\bibinfo {year} {2019})}\BibitemShut {NoStop}%
\bibitem [{\citenamefont {Bieniek}\ \emph {et~al.}(2018)\citenamefont
  {Bieniek}, \citenamefont {Korkusi\'{n}ski}, \citenamefont {Szulakowska},
  \citenamefont {Potasz}, \citenamefont {Ozfidan},\ and\ \citenamefont
  {Hawrylak}}]{atomPRB2018}%
  \BibitemOpen
  \bibfield  {author} {\bibinfo {author} {\bibfnamefont {M.}~\bibnamefont
  {Bieniek}}, \bibinfo {author} {\bibfnamefont {M.}~\bibnamefont
  {Korkusi\'{n}ski}}, \bibinfo {author} {\bibfnamefont {L.}~\bibnamefont
  {Szulakowska}}, \bibinfo {author} {\bibfnamefont {P.}~\bibnamefont {Potasz}},
  \bibinfo {author} {\bibfnamefont {I.}~\bibnamefont {Ozfidan}}, \ and\
  \bibinfo {author} {\bibfnamefont {P.}~\bibnamefont {Hawrylak}},\ }\href@noop
  {} {\bibfield  {journal} {\bibinfo  {journal} {Phys. Rev. B}\ }\textbf
  {\bibinfo {volume} {97}},\ \bibinfo {pages} {085153} (\bibinfo {year}
  {2018})}\BibitemShut {NoStop}%
\bibitem [{\citenamefont {Xiao}\ \emph {et~al.}(2007)\citenamefont {Xiao},
  \citenamefont {Yao},\ and\ \citenamefont {Niu}}]{XD2007}%
  \BibitemOpen
  \bibfield  {author} {\bibinfo {author} {\bibfnamefont {D.}~\bibnamefont
  {Xiao}}, \bibinfo {author} {\bibfnamefont {W.}~\bibnamefont {Yao}}, \ and\
  \bibinfo {author} {\bibfnamefont {Q.}~\bibnamefont {Niu}},\ }\href@noop {}
  {\bibfield  {journal} {\bibinfo  {journal} {Phys. Rev. lett.}\ }\textbf
  {\bibinfo {volume} {99}},\ \bibinfo {pages} {236809} (\bibinfo {year}
  {2007})}\BibitemShut {NoStop}%
\bibitem [{\citenamefont {Li}\ \emph {et~al.}(2014)\citenamefont {Li},
  \citenamefont {Ludwig}, \citenamefont {Low}, \citenamefont {Chernikov},
  \citenamefont {Cui}, \citenamefont {Arefe}, \citenamefont {Kim},
  \citenamefont {van~der Zande}, \citenamefont {Rigosi}, \citenamefont {Hill},
  \citenamefont {Kim}, \citenamefont {Hone}, \citenamefont {Li}, \citenamefont
  {Smirnov},\ and\ \citenamefont {Heinz}}]{PRL2014}%
  \BibitemOpen
  \bibfield  {author} {\bibinfo {author} {\bibfnamefont {Y.}~\bibnamefont
  {Li}}, \bibinfo {author} {\bibfnamefont {J.}~\bibnamefont {Ludwig}}, \bibinfo
  {author} {\bibfnamefont {T.}~\bibnamefont {Low}}, \bibinfo {author}
  {\bibfnamefont {A.}~\bibnamefont {Chernikov}}, \bibinfo {author}
  {\bibfnamefont {X.}~\bibnamefont {Cui}}, \bibinfo {author} {\bibfnamefont
  {G.}~\bibnamefont {Arefe}}, \bibinfo {author} {\bibfnamefont {Y.~D.}\
  \bibnamefont {Kim}}, \bibinfo {author} {\bibfnamefont {A.~M.}\ \bibnamefont
  {van~der Zande}}, \bibinfo {author} {\bibfnamefont {A.}~\bibnamefont
  {Rigosi}}, \bibinfo {author} {\bibfnamefont {H.~M.}\ \bibnamefont {Hill}},
  \bibinfo {author} {\bibfnamefont {S.~H.}\ \bibnamefont {Kim}}, \bibinfo
  {author} {\bibfnamefont {J.}~\bibnamefont {Hone}}, \bibinfo {author}
  {\bibfnamefont {Z.}~\bibnamefont {Li}}, \bibinfo {author} {\bibfnamefont
  {D.}~\bibnamefont {Smirnov}}, \ and\ \bibinfo {author} {\bibfnamefont
  {T.~F.}\ \bibnamefont {Heinz}},\ }\href@noop {} {\bibfield  {journal}
  {\bibinfo  {journal} {Phys. Rev. Lett.}\ }\textbf {\bibinfo {volume} {113}},\
  \bibinfo {pages} {266804} (\bibinfo {year} {2014})}\BibitemShut {NoStop}%
\bibitem [{\citenamefont {Rybkovskiy}\ \emph {et~al.}(2017)\citenamefont
  {Rybkovskiy}, \citenamefont {Gerber},\ and\ \citenamefont
  {Durnev}}]{kpPRB2017}%
  \BibitemOpen
  \bibfield  {author} {\bibinfo {author} {\bibfnamefont {D.~V.}\ \bibnamefont
  {Rybkovskiy}}, \bibinfo {author} {\bibfnamefont {I.~C.}\ \bibnamefont
  {Gerber}}, \ and\ \bibinfo {author} {\bibfnamefont {M.~V.}\ \bibnamefont
  {Durnev}},\ }\href@noop {} {\bibfield  {journal} {\bibinfo  {journal} {Phys.
  Rev. B}\ }\textbf {\bibinfo {volume} {95}},\ \bibinfo {pages} {155406}
  (\bibinfo {year} {2017})}\BibitemShut {NoStop}%
\bibitem [{\citenamefont {Korm\'{a}nyos}\ \emph {et~al.}(2015)\citenamefont
  {Korm\'{a}nyos}, \citenamefont {Rakyta},\ and\ \citenamefont
  {Burkard}}]{NewJPhys}%
  \BibitemOpen
  \bibfield  {author} {\bibinfo {author} {\bibfnamefont {A.}~\bibnamefont
  {Korm\'{a}nyos}}, \bibinfo {author} {\bibfnamefont {P.}~\bibnamefont
  {Rakyta}}, \ and\ \bibinfo {author} {\bibfnamefont {G.}~\bibnamefont
  {Burkard}},\ }\href@noop {} {\bibfield  {journal} {\bibinfo  {journal} {New
  J. Phys.}\ }\textbf {\bibinfo {volume} {17}},\ \bibinfo {pages} {103006}
  (\bibinfo {year} {2015})}\BibitemShut {NoStop}%
\bibitem [{\citenamefont {Luttinger}\ and\ \citenamefont
  {Kohn}(1955)}]{LK1955}%
  \BibitemOpen
  \bibfield  {author} {\bibinfo {author} {\bibfnamefont {J.~M.}\ \bibnamefont
  {Luttinger}}\ and\ \bibinfo {author} {\bibfnamefont {W.}~\bibnamefont
  {Kohn}},\ }\href@noop {} {\bibfield  {journal} {\bibinfo  {journal} {Phys.
  Rev.}\ }\textbf {\bibinfo {volume} {97}},\ \bibinfo {pages} {869} (\bibinfo
  {year} {1955})}\BibitemShut {NoStop}%
\bibitem [{\citenamefont {Luttinger}(1951)}]{Lut1951}%
  \BibitemOpen
  \bibfield  {author} {\bibinfo {author} {\bibfnamefont {J.~M.}\ \bibnamefont
  {Luttinger}},\ }\href@noop {} {\bibfield  {journal} {\bibinfo  {journal}
  {Phys. Rev.}\ }\textbf {\bibinfo {volume} {84}},\ \bibinfo {pages} {814}
  (\bibinfo {year} {1951})}\BibitemShut {NoStop}%
\bibitem [{\citenamefont {Liu}\ \emph {et~al.}(2019{\natexlab{a}})\citenamefont
  {Liu}, \citenamefont {van Baren}, \citenamefont {Taniguchi}, \citenamefont
  {Watanabe}, \citenamefont {Chang},\ and\ \citenamefont {Lui}}]{PRR2019}%
  \BibitemOpen
  \bibfield  {author} {\bibinfo {author} {\bibfnamefont {E.}~\bibnamefont
  {Liu}}, \bibinfo {author} {\bibfnamefont {J.}~\bibnamefont {van Baren}},
  \bibinfo {author} {\bibfnamefont {T.}~\bibnamefont {Taniguchi}}, \bibinfo
  {author} {\bibfnamefont {K.}~\bibnamefont {Watanabe}}, \bibinfo {author}
  {\bibfnamefont {Y.-C.}\ \bibnamefont {Chang}}, \ and\ \bibinfo {author}
  {\bibfnamefont {C.~H.}\ \bibnamefont {Lui}},\ }\href@noop {} {\bibfield
  {journal} {\bibinfo  {journal} {Phys. Rev. Res.}\ }\textbf {\bibinfo {volume}
  {1}},\ \bibinfo {pages} {032007(R)} (\bibinfo {year}
  {2019}{\natexlab{a}})}\BibitemShut {NoStop}%
\bibitem [{\citenamefont {Hofstadter}(1976)}]{Hof}%
  \BibitemOpen
  \bibfield  {author} {\bibinfo {author} {\bibfnamefont {D.~R.}\ \bibnamefont
  {Hofstadter}},\ }\href@noop {} {\bibfield  {journal} {\bibinfo  {journal}
  {Phys. Rev. B}\ }\textbf {\bibinfo {volume} {14}},\ \bibinfo {pages} {2239}
  (\bibinfo {year} {1976})}\BibitemShut {NoStop}%
\bibitem [{\citenamefont {Ceresoli}\ \emph {et~al.}(2006)\citenamefont
  {Ceresoli}, \citenamefont {Thonharser}, \citenamefont {Vanderbilt},\ and\
  \citenamefont {Resta}}]{VanPRB}%
  \BibitemOpen
  \bibfield  {author} {\bibinfo {author} {\bibfnamefont {D.}~\bibnamefont
  {Ceresoli}}, \bibinfo {author} {\bibfnamefont {T.}~\bibnamefont
  {Thonharser}}, \bibinfo {author} {\bibfnamefont {D.}~\bibnamefont
  {Vanderbilt}}, \ and\ \bibinfo {author} {\bibfnamefont {R.}~\bibnamefont
  {Resta}},\ }\href@noop {} {\bibfield  {journal} {\bibinfo  {journal} {Phys.
  Rev. B}\ }\textbf {\bibinfo {volume} {74}},\ \bibinfo {pages} {024408}
  (\bibinfo {year} {2006})}\BibitemShut {NoStop}%
\bibitem [{\citenamefont {Giannozzi}\ \emph {et~al.}(2009)\citenamefont
  {Giannozzi}, \citenamefont {Baroni}, \citenamefont {Bonini}, \citenamefont
  {Calandra}, \citenamefont {Car}, \citenamefont {Cavazzoni}, \citenamefont
  {Ceresoli}, \citenamefont {Chiarotti}, \citenamefont {Cococcioni},
  \citenamefont {Dabo},\ and\ \citenamefont {et~al.}}]{QE}%
  \BibitemOpen
  \bibfield  {author} {\bibinfo {author} {\bibfnamefont {P.}~\bibnamefont
  {Giannozzi}}, \bibinfo {author} {\bibfnamefont {S.}~\bibnamefont {Baroni}},
  \bibinfo {author} {\bibfnamefont {N.}~\bibnamefont {Bonini}}, \bibinfo
  {author} {\bibfnamefont {M.}~\bibnamefont {Calandra}}, \bibinfo {author}
  {\bibfnamefont {R.}~\bibnamefont {Car}}, \bibinfo {author} {\bibfnamefont
  {C.}~\bibnamefont {Cavazzoni}}, \bibinfo {author} {\bibfnamefont
  {D.}~\bibnamefont {Ceresoli}}, \bibinfo {author} {\bibfnamefont {G.~L.}\
  \bibnamefont {Chiarotti}}, \bibinfo {author} {\bibfnamefont {M.}~\bibnamefont
  {Cococcioni}}, \bibinfo {author} {\bibfnamefont {I.}~\bibnamefont {Dabo}}, \
  and\ \bibinfo {author} {\bibnamefont {et~al.}},\ }\href@noop {} {\bibfield
  {journal} {\bibinfo  {journal} {J. Phys.: Condens. Matter}\ }\textbf
  {\bibinfo {volume} {21}},\ \bibinfo {pages} {395502} (\bibinfo {year}
  {2009})}\BibitemShut {NoStop}%
\bibitem [{\citenamefont {Perdew}\ \emph {et~al.}(1996)\citenamefont {Perdew},
  \citenamefont {Burke},\ and\ \citenamefont {Ernzerhof}}]{PBE1996}%
  \BibitemOpen
  \bibfield  {author} {\bibinfo {author} {\bibfnamefont {J.~P.}\ \bibnamefont
  {Perdew}}, \bibinfo {author} {\bibfnamefont {K.}~\bibnamefont {Burke}}, \
  and\ \bibinfo {author} {\bibfnamefont {M.}~\bibnamefont {Ernzerhof}},\
  }\href@noop {} {\bibfield  {journal} {\bibinfo  {journal} {Phys. Rev. Lett.}\
  }\textbf {\bibinfo {volume} {77}},\ \bibinfo {pages} {3865} (\bibinfo {year}
  {1996})}\BibitemShut {NoStop}%
\bibitem [{\citenamefont {Vignale}\ and\ \citenamefont
  {Rasolt}(1987)}]{PRL1987}%
  \BibitemOpen
  \bibfield  {author} {\bibinfo {author} {\bibfnamefont {G.}~\bibnamefont
  {Vignale}}\ and\ \bibinfo {author} {\bibfnamefont {M.}~\bibnamefont
  {Rasolt}},\ }\href@noop {} {\bibfield  {journal} {\bibinfo  {journal} {Phys.
  Rev. Lett.}\ }\textbf {\bibinfo {volume} {59}},\ \bibinfo {pages} {2360}
  (\bibinfo {year} {1987})}\BibitemShut {NoStop}%
\bibitem [{\citenamefont {Mauri}\ \emph {et~al.}(1996)\citenamefont {Mauri},
  \citenamefont {Pfrommer},\ and\ \citenamefont {Louie}}]{NMR1996}%
  \BibitemOpen
  \bibfield  {author} {\bibinfo {author} {\bibfnamefont {F.}~\bibnamefont
  {Mauri}}, \bibinfo {author} {\bibfnamefont {B.~G.}\ \bibnamefont {Pfrommer}},
  \ and\ \bibinfo {author} {\bibfnamefont {S.~G.}\ \bibnamefont {Louie}},\
  }\href@noop {} {\bibfield  {journal} {\bibinfo  {journal} {Phys. Rev. Lett.}\
  }\textbf {\bibinfo {volume} {77}},\ \bibinfo {pages} {5300} (\bibinfo {year}
  {1996})}\BibitemShut {NoStop}%
\bibitem [{\citenamefont {Schlipf}\ and\ \citenamefont {Gygi}(2015)}]{oncv}%
  \BibitemOpen
  \bibfield  {author} {\bibinfo {author} {\bibfnamefont {M.}~\bibnamefont
  {Schlipf}}\ and\ \bibinfo {author} {\bibfnamefont {F.}~\bibnamefont {Gygi}},\
  }\href@noop {} {\bibfield  {journal} {\bibinfo  {journal} {Comput. Phys.
  Commun.}\ }\textbf {\bibinfo {volume} {196}},\ \bibinfo {pages} {36}
  (\bibinfo {year} {2015})}\BibitemShut {NoStop}%
\bibitem [{\citenamefont {Hybertsen}\ and\ \citenamefont
  {Louie}(1986)}]{Louie1986}%
  \BibitemOpen
  \bibfield  {author} {\bibinfo {author} {\bibfnamefont {M.~S.}\ \bibnamefont
  {Hybertsen}}\ and\ \bibinfo {author} {\bibfnamefont {S.~G.}\ \bibnamefont
  {Louie}},\ }\href@noop {} {\bibfield  {journal} {\bibinfo  {journal} {Phys.
  Rev. B}\ }\textbf {\bibinfo {volume} {34}},\ \bibinfo {pages} {5390}
  (\bibinfo {year} {1986})}\BibitemShut {NoStop}%
\bibitem [{\citenamefont {Qiu}\ \emph {et~al.}(2013)\citenamefont {Qiu},
  \citenamefont {da~Jornada},\ and\ \citenamefont {Louie}}]{Louie2013PRL}%
  \BibitemOpen
  \bibfield  {author} {\bibinfo {author} {\bibfnamefont {D.~Y.}\ \bibnamefont
  {Qiu}}, \bibinfo {author} {\bibfnamefont {F.~H.}\ \bibnamefont {da~Jornada}},
  \ and\ \bibinfo {author} {\bibfnamefont {S.~G.}\ \bibnamefont {Louie}},\
  }\href@noop {} {\bibfield  {journal} {\bibinfo  {journal} {Phys. Rev. Lett.}\
  }\textbf {\bibinfo {volume} {111}},\ \bibinfo {pages} {216805} (\bibinfo
  {year} {2013})}\BibitemShut {NoStop}%
\bibitem [{\citenamefont {Deslippe}\ \emph {et~al.}(2012)\citenamefont
  {Deslippe}, \citenamefont {Samsonidze}, \citenamefont {Strubble},
  \citenamefont {Jain}, \citenamefont {Cohen},\ and\ \citenamefont
  {Louie}}]{BGW}%
  \BibitemOpen
  \bibfield  {author} {\bibinfo {author} {\bibfnamefont {J.}~\bibnamefont
  {Deslippe}}, \bibinfo {author} {\bibfnamefont {G.}~\bibnamefont
  {Samsonidze}}, \bibinfo {author} {\bibfnamefont {D.~A.}\ \bibnamefont
  {Strubble}}, \bibinfo {author} {\bibfnamefont {M.}~\bibnamefont {Jain}},
  \bibinfo {author} {\bibfnamefont {M.~L.}\ \bibnamefont {Cohen}}, \ and\
  \bibinfo {author} {\bibfnamefont {S.~G.}\ \bibnamefont {Louie}},\ }\href@noop
  {} {\bibfield  {journal} {\bibinfo  {journal} {Comput. Phys. Commun.}\
  }\textbf {\bibinfo {volume} {183}},\ \bibinfo {pages} {1269} (\bibinfo {year}
  {2012})}\BibitemShut {NoStop}%
\bibitem [{\citenamefont {Hedin}(1965)}]{Hedin1965}%
  \BibitemOpen
  \bibfield  {author} {\bibinfo {author} {\bibfnamefont {L.}~\bibnamefont
  {Hedin}},\ }\href@noop {} {\bibfield  {journal} {\bibinfo  {journal} {Phys.
  Rev.}\ }\textbf {\bibinfo {volume} {139}},\ \bibinfo {pages} {A796} (\bibinfo
  {year} {1965})}\BibitemShut {NoStop}%
\bibitem [{\citenamefont {Qiu}\ \emph {et~al.}(2016)\citenamefont {Qiu},
  \citenamefont {da~Jornada},\ and\ \citenamefont {Louie}}]{Qiu2016}%
  \BibitemOpen
  \bibfield  {author} {\bibinfo {author} {\bibfnamefont {D.~Y.}\ \bibnamefont
  {Qiu}}, \bibinfo {author} {\bibfnamefont {F.~H.}\ \bibnamefont {da~Jornada}},
  \ and\ \bibinfo {author} {\bibfnamefont {S.~G.}\ \bibnamefont {Louie}},\
  }\href@noop {} {\bibfield  {journal} {\bibinfo  {journal} {Phys. Rev. B}\
  }\textbf {\bibinfo {volume} {93}},\ \bibinfo {pages} {235435} (\bibinfo
  {year} {2016})}\BibitemShut {NoStop}%
\bibitem [{\citenamefont {Jornada}\ \emph {et~al.}(2017)\citenamefont
  {Jornada}, , \citenamefont {Qiu},\ and\ \citenamefont {Louie}}]{sub}%
  \BibitemOpen
  \bibfield  {author} {\bibinfo {author} {\bibfnamefont {F.~d.}\ \bibnamefont
  {Jornada}}, , \bibinfo {author} {\bibfnamefont {D.~Y.}\ \bibnamefont {Qiu}},
  \ and\ \bibinfo {author} {\bibfnamefont {S.~G.}\ \bibnamefont {Louie}},\
  }\href@noop {} {\bibfield  {journal} {\bibinfo  {journal} {Phys. Rev. B}\
  }\textbf {\bibinfo {volume} {95}},\ \bibinfo {pages} {035109} (\bibinfo
  {year} {2017})}\BibitemShut {NoStop}%
\bibitem [{\citenamefont {Liu}\ \emph {et~al.}(2013)\citenamefont {Liu},
  \citenamefont {Shan}, \citenamefont {Yao}, \citenamefont {Yao},\ and\
  \citenamefont {Xiao}}]{TB2013}%
  \BibitemOpen
  \bibfield  {author} {\bibinfo {author} {\bibfnamefont {G.-B.}\ \bibnamefont
  {Liu}}, \bibinfo {author} {\bibfnamefont {W.-Y.}\ \bibnamefont {Shan}},
  \bibinfo {author} {\bibfnamefont {Y.}~\bibnamefont {Yao}}, \bibinfo {author}
  {\bibfnamefont {W.}~\bibnamefont {Yao}}, \ and\ \bibinfo {author}
  {\bibfnamefont {D.}~\bibnamefont {Xiao}},\ }\href@noop {} {\bibfield
  {journal} {\bibinfo  {journal} {Phys. Rev. B}\ }\textbf {\bibinfo {volume}
  {88}},\ \bibinfo {pages} {085433} (\bibinfo {year} {2013})}\BibitemShut
  {NoStop}%
\bibitem [{\citenamefont {Feng}\ \emph {et~al.}(2012)\citenamefont {Feng},
  \citenamefont {Yao}, \citenamefont {Zhu}, \citenamefont {Zhou}, \citenamefont
  {Yao},\ and\ \citenamefont {Xiao}}]{XD2012}%
  \BibitemOpen
  \bibfield  {author} {\bibinfo {author} {\bibfnamefont {W.}~\bibnamefont
  {Feng}}, \bibinfo {author} {\bibfnamefont {Y.}~\bibnamefont {Yao}}, \bibinfo
  {author} {\bibfnamefont {W.}~\bibnamefont {Zhu}}, \bibinfo {author}
  {\bibfnamefont {J.}~\bibnamefont {Zhou}}, \bibinfo {author} {\bibfnamefont
  {W.}~\bibnamefont {Yao}}, \ and\ \bibinfo {author} {\bibfnamefont
  {D.}~\bibnamefont {Xiao}},\ }\href@noop {} {\bibfield  {journal} {\bibinfo
  {journal} {Phys. Rev. B}\ }\textbf {\bibinfo {volume} {86}},\ \bibinfo
  {pages} {165108} (\bibinfo {year} {2012})}\BibitemShut {NoStop}%
\bibitem [{\citenamefont {Fang}\ \emph {et~al.}(2015)\citenamefont {Fang},
  \citenamefont {Defo}, \citenamefont {Shirodkar}, \citenamefont {Lieu},
  \citenamefont {Tritsaris},\ and\ \citenamefont {Kaxiras}}]{TB11}%
  \BibitemOpen
  \bibfield  {author} {\bibinfo {author} {\bibfnamefont {S.}~\bibnamefont
  {Fang}}, \bibinfo {author} {\bibfnamefont {R.~K.}\ \bibnamefont {Defo}},
  \bibinfo {author} {\bibfnamefont {S.~N.}\ \bibnamefont {Shirodkar}}, \bibinfo
  {author} {\bibfnamefont {S.}~\bibnamefont {Lieu}}, \bibinfo {author}
  {\bibfnamefont {A.}~\bibnamefont {Tritsaris}}, \ and\ \bibinfo {author}
  {\bibfnamefont {E.}~\bibnamefont {Kaxiras}},\ }\href@noop {} {\bibfield
  {journal} {\bibinfo  {journal} {Phys. Rev. B}\ }\textbf {\bibinfo {volume}
  {92}},\ \bibinfo {pages} {205108} (\bibinfo {year} {2015})}\BibitemShut
  {NoStop}%
\bibitem [{\citenamefont {Solovyev}\ and\ \citenamefont
  {Nikolaev}(2014)}]{solo2014}%
  \BibitemOpen
  \bibfield  {author} {\bibinfo {author} {\bibfnamefont {I.~V.}\ \bibnamefont
  {Solovyev}}\ and\ \bibinfo {author} {\bibfnamefont {S.~A.}\ \bibnamefont
  {Nikolaev}},\ }\href@noop {} {\bibfield  {journal} {\bibinfo  {journal}
  {Phys. Rev. B}\ }\textbf {\bibinfo {volume} {89}},\ \bibinfo {pages} {064428}
  (\bibinfo {year} {2014})}\BibitemShut {NoStop}%
\bibitem [{\citenamefont {Gradhand}\ \emph {et~al.}(2012)\citenamefont
  {Gradhand}, \citenamefont {Fedorov}, \citenamefont {Pientka}, \citenamefont
  {Zahn}, \citenamefont {Mertig},\ and\ \citenamefont
  {Gy\"{o}rffy}}]{JPCM2012}%
  \BibitemOpen
  \bibfield  {author} {\bibinfo {author} {\bibfnamefont {M.}~\bibnamefont
  {Gradhand}}, \bibinfo {author} {\bibfnamefont {D.~V.}\ \bibnamefont
  {Fedorov}}, \bibinfo {author} {\bibfnamefont {F.}~\bibnamefont {Pientka}},
  \bibinfo {author} {\bibfnamefont {P.}~\bibnamefont {Zahn}}, \bibinfo {author}
  {\bibfnamefont {I.}~\bibnamefont {Mertig}}, \ and\ \bibinfo {author}
  {\bibfnamefont {B.~L.}\ \bibnamefont {Gy\"{o}rffy}},\ }\href@noop {}
  {\bibfield  {journal} {\bibinfo  {journal} {J. Phys.: Condens. Matter}\
  }\textbf {\bibinfo {volume} {24}},\ \bibinfo {pages} {213202} (\bibinfo
  {year} {2012})}\BibitemShut {NoStop}%
\bibitem [{\citenamefont {Li}\ \emph {et~al.}(2019)\citenamefont {Li},
  \citenamefont {Wang}, \citenamefont {Jin}, \citenamefont {Lu}, \citenamefont
  {Lian}, \citenamefont {Meng}, \citenamefont {Blei}, \citenamefont {Gao},
  \citenamefont {Taniguchi}, \citenamefont {Watanabe}, \citenamefont {Ren},
  \citenamefont {Tongay}, \citenamefont {Yang}, \citenamefont {Smirnov},
  \citenamefont {Cao},\ and\ \citenamefont {Shi}}]{NC2019}%
  \BibitemOpen
  \bibfield  {author} {\bibinfo {author} {\bibfnamefont {Z.}~\bibnamefont
  {Li}}, \bibinfo {author} {\bibfnamefont {T.}~\bibnamefont {Wang}}, \bibinfo
  {author} {\bibfnamefont {C.}~\bibnamefont {Jin}}, \bibinfo {author}
  {\bibfnamefont {Z.}~\bibnamefont {Lu}}, \bibinfo {author} {\bibfnamefont
  {Z.}~\bibnamefont {Lian}}, \bibinfo {author} {\bibfnamefont {Y.}~\bibnamefont
  {Meng}}, \bibinfo {author} {\bibfnamefont {M.}~\bibnamefont {Blei}}, \bibinfo
  {author} {\bibfnamefont {S.}~\bibnamefont {Gao}}, \bibinfo {author}
  {\bibfnamefont {T.}~\bibnamefont {Taniguchi}}, \bibinfo {author}
  {\bibfnamefont {K.}~\bibnamefont {Watanabe}}, \bibinfo {author}
  {\bibfnamefont {T.}~\bibnamefont {Ren}}, \bibinfo {author} {\bibfnamefont
  {S.}~\bibnamefont {Tongay}}, \bibinfo {author} {\bibfnamefont
  {L.}~\bibnamefont {Yang}}, \bibinfo {author} {\bibfnamefont {D.}~\bibnamefont
  {Smirnov}}, \bibinfo {author} {\bibfnamefont {T.}~\bibnamefont {Cao}}, \ and\
  \bibinfo {author} {\bibfnamefont {S.-F.}\ \bibnamefont {Shi}},\ }\href@noop
  {} {\bibfield  {journal} {\bibinfo  {journal} {Nat. Commun.}\ }\textbf
  {\bibinfo {volume} {10}},\ \bibinfo {pages} {2469} (\bibinfo {year}
  {2019})}\BibitemShut {NoStop}%
\bibitem [{\citenamefont {Liu}\ \emph {et~al.}(2019{\natexlab{b}})\citenamefont
  {Liu}, \citenamefont {van Baren}, \citenamefont {Lu}, \citenamefont
  {Altaiary}, \citenamefont {Taniguchi}, \citenamefont {Watanabe},
  \citenamefont {Smirnov},\ and\ \citenamefont {Lui}}]{Liu2019}%
  \BibitemOpen
  \bibfield  {author} {\bibinfo {author} {\bibfnamefont {E.}~\bibnamefont
  {Liu}}, \bibinfo {author} {\bibfnamefont {J.}~\bibnamefont {van Baren}},
  \bibinfo {author} {\bibfnamefont {Z.}~\bibnamefont {Lu}}, \bibinfo {author}
  {\bibfnamefont {M.~M.}\ \bibnamefont {Altaiary}}, \bibinfo {author}
  {\bibfnamefont {T.}~\bibnamefont {Taniguchi}}, \bibinfo {author}
  {\bibfnamefont {K.}~\bibnamefont {Watanabe}}, \bibinfo {author}
  {\bibfnamefont {D.}~\bibnamefont {Smirnov}}, \ and\ \bibinfo {author}
  {\bibfnamefont {C.~H.}\ \bibnamefont {Lui}},\ }\href@noop {} {\bibfield
  {journal} {\bibinfo  {journal} {Phys. Rev. Lett.}\ }\textbf {\bibinfo
  {volume} {123}},\ \bibinfo {pages} {027401} (\bibinfo {year}
  {2019}{\natexlab{b}})}\BibitemShut {NoStop}%
\bibitem [{\citenamefont {Lebedeva}\ \emph {et~al.}(2019)\citenamefont
  {Lebedeva}, \citenamefont {Strubble}, \citenamefont {Tokatly},\ and\
  \citenamefont {Rubio}}]{Rubio}%
  \BibitemOpen
  \bibfield  {author} {\bibinfo {author} {\bibfnamefont {I.~V.}\ \bibnamefont
  {Lebedeva}}, \bibinfo {author} {\bibfnamefont {D.}~\bibnamefont {Strubble}},
  \bibinfo {author} {\bibfnamefont {I.~V.}\ \bibnamefont {Tokatly}}, \ and\
  \bibinfo {author} {\bibfnamefont {A.}~\bibnamefont {Rubio}},\ }\href@noop {}
  {\bibfield  {journal} {\bibinfo  {journal} {NPJ Comput. Mater.}\ }\textbf
  {\bibinfo {volume} {5}},\ \bibinfo {pages} {32} (\bibinfo {year}
  {2019})}\BibitemShut {NoStop}%
\bibitem [{\citenamefont {Xuan}\ and\ \citenamefont {Quek}(shed)}]{pre}%
  \BibitemOpen
  \bibfield  {author} {\bibinfo {author} {\bibfnamefont {F.}~\bibnamefont
  {Xuan}}\ and\ \bibinfo {author} {\bibfnamefont {S.~Y.}\ \bibnamefont
  {Quek}},\ }\href@noop {} {} (\bibinfo {year} {unpublished})\BibitemShut
  {NoStop}%
\bibitem [{\citenamefont {Roth}\ \emph {et~al.}(1959)\citenamefont {Roth},
  \citenamefont {Lax},\ and\ \citenamefont {Zwerdling}}]{Roth1959}%
  \BibitemOpen
  \bibfield  {author} {\bibinfo {author} {\bibfnamefont {L.~M.}\ \bibnamefont
  {Roth}}, \bibinfo {author} {\bibfnamefont {B.}~\bibnamefont {Lax}}, \ and\
  \bibinfo {author} {\bibfnamefont {S.}~\bibnamefont {Zwerdling}},\ }\href@noop
  {} {\bibfield  {journal} {\bibinfo  {journal} {Phys. Rev.}\ }\textbf
  {\bibinfo {volume} {114}},\ \bibinfo {pages} {90} (\bibinfo {year}
  {1959})}\BibitemShut {NoStop}%
\bibitem [{\citenamefont {Seyler}\ \emph {et~al.}(2019)\citenamefont {Seyler},
  \citenamefont {Rivera}, \citenamefont {Yu}, \citenamefont {Wilson},
  \citenamefont {Ray}, \citenamefont {Mandrus}, \citenamefont {Yan},
  \citenamefont {Yao},\ and\ \citenamefont {Xu}}]{Xu2019}%
  \BibitemOpen
  \bibfield  {author} {\bibinfo {author} {\bibfnamefont {K.~L.}\ \bibnamefont
  {Seyler}}, \bibinfo {author} {\bibfnamefont {P.}~\bibnamefont {Rivera}},
  \bibinfo {author} {\bibfnamefont {H.}~\bibnamefont {Yu}}, \bibinfo {author}
  {\bibfnamefont {N.~P.}\ \bibnamefont {Wilson}}, \bibinfo {author}
  {\bibfnamefont {E.~L.}\ \bibnamefont {Ray}}, \bibinfo {author} {\bibfnamefont
  {D.~G.}\ \bibnamefont {Mandrus}}, \bibinfo {author} {\bibfnamefont
  {J.}~\bibnamefont {Yan}}, \bibinfo {author} {\bibfnamefont {W.}~\bibnamefont
  {Yao}}, \ and\ \bibinfo {author} {\bibfnamefont {X.}~\bibnamefont {Xu}},\
  }\href@noop {} {\bibfield  {journal} {\bibinfo  {journal} {Nature}\ }\textbf
  {\bibinfo {volume} {567}},\ \bibinfo {pages} {66} (\bibinfo {year}
  {2019})}\BibitemShut {NoStop}%
\bibitem [{\citenamefont {Nagler}\ \emph {et~al.}(2017)\citenamefont {Nagler},
  \citenamefont {Ballottin}, \citenamefont {Mitioglu}, \citenamefont
  {Mooshammer}, \citenamefont {Paradiso}, \citenamefont {Strunk}, \citenamefont
  {Huber}, \citenamefont {Chernikov}, \citenamefont {Christianen},
  \citenamefont {Sch\"{u}ller},\ and\ \citenamefont {Korn}}]{NC2017}%
  \BibitemOpen
  \bibfield  {author} {\bibinfo {author} {\bibfnamefont {P.}~\bibnamefont
  {Nagler}}, \bibinfo {author} {\bibfnamefont {M.~V.}\ \bibnamefont
  {Ballottin}}, \bibinfo {author} {\bibfnamefont {A.~A.}\ \bibnamefont
  {Mitioglu}}, \bibinfo {author} {\bibfnamefont {F.}~\bibnamefont
  {Mooshammer}}, \bibinfo {author} {\bibfnamefont {N.}~\bibnamefont
  {Paradiso}}, \bibinfo {author} {\bibfnamefont {C.}~\bibnamefont {Strunk}},
  \bibinfo {author} {\bibfnamefont {R.}~\bibnamefont {Huber}}, \bibinfo
  {author} {\bibfnamefont {A.}~\bibnamefont {Chernikov}}, \bibinfo {author}
  {\bibfnamefont {P.~C.}\ \bibnamefont {Christianen}}, \bibinfo {author}
  {\bibfnamefont {C.}~\bibnamefont {Sch\"{u}ller}}, \ and\ \bibinfo {author}
  {\bibfnamefont {T.}~\bibnamefont {Korn}},\ }\href@noop {} {\bibfield
  {journal} {\bibinfo  {journal} {Nat. Commun.}\ }\textbf {\bibinfo {volume}
  {8}},\ \bibinfo {pages} {1551} (\bibinfo {year} {2017})}\BibitemShut
  {NoStop}%
\bibitem [{\citenamefont {Wo\'{z}niak}\ \emph {et~al.}()\citenamefont
  {Wo\'{z}niak}, \citenamefont {Faria~Junior}, \citenamefont {Seifert},
  \citenamefont {Chaves},\ and\ \citenamefont {Kunstmann}}]{arXiv2020}%
  \BibitemOpen
  \bibfield  {author} {\bibinfo {author} {\bibfnamefont {T.}~\bibnamefont
  {Wo\'{z}niak}}, \bibinfo {author} {\bibfnamefont {P.~E.}\ \bibnamefont
  {Faria~Junior}}, \bibinfo {author} {\bibfnamefont {G.}~\bibnamefont
  {Seifert}}, \bibinfo {author} {\bibfnamefont {A.}~\bibnamefont {Chaves}}, \
  and\ \bibinfo {author} {\bibfnamefont {J.}~\bibnamefont {Kunstmann}},\
  }\href@noop {} {\bibinfo  {journal} {arXiv:2002.02542}\ }\BibitemShut
  {NoStop}%
\bibitem [{\citenamefont {Deilmann}\ \emph {et~al.}()\citenamefont {Deilmann},
  \citenamefont {Kr\"{u}ger},\ and\ \citenamefont {Rohlfing}}]{arXivBSE}%
  \BibitemOpen
\bibfield  {journal} {  }\bibfield  {author} {\bibinfo {author} {\bibfnamefont
  {T.}~\bibnamefont {Deilmann}}, \bibinfo {author} {\bibfnamefont
  {P.}~\bibnamefont {Kr\"{u}ger}}, \ and\ \bibinfo {author} {\bibfnamefont
  {M.}~\bibnamefont {Rohlfing}},\ }\href@noop {} {\bibinfo  {journal}
  {arXiv:2003.00235}\ }\BibitemShut {NoStop}%
\bibitem [{\citenamefont {Mott}\ and\ \citenamefont {Massey}(1965)}]{Mott}%
  \BibitemOpen
\bibfield  {journal} {  }\bibfield  {author} {\bibinfo {author} {\bibfnamefont
  {N.~F.}\ \bibnamefont {Mott}}\ and\ \bibinfo {author} {\bibfnamefont
  {H.~S.~W.}\ \bibnamefont {Massey}},\ }\href@noop {} {\emph {\bibinfo {title}
  {The theory of atomic collisions.}}}\ (\bibinfo  {publisher} {Oxford
  University Press},\ \bibinfo {year} {1965})\ p.\ \bibinfo {pages}
  {181}\BibitemShut {NoStop}%
\end{thebibliography}%

\end{document}